\documentclass{aa}

\usepackage[varg]{txfonts}
\usepackage{csquotes}
\usepackage{hyperref}
\usepackage{booktabs}
\usepackage{amsmath, amssymb, bbold, mathrsfs, nicefrac, dsfont}
\usepackage{graphicx}
\graphicspath{{img/}}

\newcommand{\prob}{\mathcal P }

\newcommand{\intds}{\int \mathcal Ds\,}
\newcommand{\intdn}{\int \mathcal Dn\,}
\newcommand{\post}{\prob (s|d)}
\newcommand{\lh}{\prob (d|s)}
\newcommand{\prior}{\prob (s)}
\newcommand{\gauss}[2]{\mathcal G (#1, #2)}
\newcommand{\gaussauto}[2]{\mathcal G \left(#1, #2\right)}

\begin{document}

\title{Unified radio interferometric calibration and imaging\\with joint uncertainty quantification} %
\author{ Philipp Arras\inst{\ref{mpa}\ref{lmu}\ref{tum}}\and Philipp Frank\inst{\ref{mpa}\ref{lmu}}\and Reimar Leike\inst{\ref{mpa}\ref{lmu}}\and Rüdiger Westermann\inst{\ref{tum}}\and Torsten A. Enßlin\inst{\ref{mpa}\ref{lmu}}} %
\institute{ Max-Planck Institut für Astrophysik, Karl-Schwarzschild-Str. 1, Garching, Germany\label{mpa}\and Ludwig-Maximilians-Universität München (LMU), Geschwister-Scholl-Platz 1, München, Germany\label{lmu}\and Technische Universität München (TUM), Boltzmannstr. 3, 85748 Garching, Germany\label{tum} }

\date{Received <date>/ Accepted <date>}

\abstract{The data reduction procedure for radio interferometers can be viewed as a combined calibration and imaging problem. %
  We present an algorithm that unifies cross-calibration, self-calibration, and imaging. %
  Because it is a Bayesian method, this algorithm not only calculates an estimate of the sky brightness distribution, but also provides an estimate of the joint uncertainty which entails both the uncertainty of the calibration and that of the actual observation. %
  The algorithm is formulated in the language of information field theory and uses Metric Gaussian Variational Inference (MGVI) as the underlying statistical method. %
  So far only direction-independent antenna-based calibration is considered. %
  This restriction may be released in future work. %
  An implementation of the algorithm is contributed as well.} \keywords{techniques: interferometric -- methods: statistical -- methods: data analysis -- instrumentation: interferometers}
\maketitle


\section{Introduction}
Radio astronomy is thriving. %
Super-modern telescopes such as MeerKAT, the Australian Square Kilometer Array Pathfinder (ASKAP), the Very Large Array (VLA), and the Atacama Large Millimeter Array (ALMA) are operating and the Square Kilometer Array (SKA) is in the planning stages. %
All these telescopes provide high-quality data on an unprecedented scale and much progress is being made instrumental-wise, which facilitates enormous improvements in sensitivity and survey speed. %

Impressed by these novel facilities we would like to turn our attention to the calibration and imaging algorithms that are fed by the data from these telescopes. %
The amount of scientific insight that can possibly be extracted from a given telescope is limited by the capability of the employed data reduction algorithm. %
We suggest that there is room for improvement regarding the calibration and imaging procedure: the most widely applied algorithms view calibration and imaging as separate problems and are not able to provide uncertainty information. %
The latter is desperately needed to quantify the level of trust a scientist can put on any result based on radio observations. %
Furthermore, a statistical sound confrontation of astrophysical models to radio data requires reliable uncertainty quantification. %
Treating calibration and imaging as separate steps ignores their tight interdependence. %

The algorithmic idea presented in this work is an advancement of the original \textsc{resolve} algorithm (\textbf{R}adio \textbf{E}xtended \textbf{SO}urces \textbf{L}ognormal decon\textbf{v}olution \textbf{E}stimator; \citealt{Resolve2016}) and may retain its name. %
The \textsc{resolve} algorithm is formulated in the language of information field theory (IFT; \citealt{ensslin2009information, ensslin2018information}), which is a view on Bayesian statistics applicable wherever (physical) fields are supposed to be inferred. %
From a Bayesian point of view the question when reducing radio data is the following: Given prior knowledge as well as measurement information about the brightness distribution of a patch of the sky, what knowledge does the observer have after obtaining the data? %
This question is answered by the Bayes theorem in terms of a probability distribution over all possible sky brightness distributions conditional to the data.

Reconstruction algorithms may be judged based on their statistical integrity or by their performance. %
The first perspective ultimately leads to pure Bayesian algorithms, which are too expensive for typical problems computationally. %
The latter often leads to ad hoc algorithms that may perform well in applications, but these can have major shortcomings such as a missing uncertainty quantification or negative-flux pixels, which is the case, for example, for \textsc{CLEAN} \citep{hogbom1974aperture}. %
The \textsc{resolve} algorithm attempts a compromise between these two objectives. %
It is based on purely statistical arguments and the necessary operations are approximated such that they can efficiently be implemented on a computer and be used for actual imaging tasks. %
Thus, the approximations and (prior) assumptions on which \textsc{resolve} is based can be written down explicitly.

\textsc{resolve} is reasonably fast but cannot compete in pure speed with algorithms like the Cotton-Schwab algorithm \citep{schwab1984relaxing} as implemented in CASA. %
This is rooted in the fact that \textsc{resolve} not only provides a single sky brightness distribution but needs to update the sky prior probability distribution according to the raw data in order to properly state how much the data has constrained the probability distribution and how much uncertainty is left in the final result. %
This uncertainty is defined in a fashion such that it can encode the posterior variance and also cross-correlations. %
Thus, the uncertainty is qunatified by $\mathcal O (n^2)$ pieces of information where $n$ is the number of pixels in the image. %
Given this massive amount of degrees of freedom it may be surprising that \textsc{resolve} is able to return its results after a sensible amount of time. %
Having said this, there is still potential for improvement. %
The technical cause for the long runtime is the complexity of the gridding/degridding operation, which needs to be called orders of magnitude more often than for conventional algorithms. %
This problem may be tackled from an information-theoretic perspective in the future.

Turning to the specific subject of the present publication, the data reduction pipeline of modern radio telescopes consists of numerous steps. %
In this paper, we would like to focus on the calibration and imaging part. %
Calibration is necessary because the data is corrupted by a variety of effects including antenna-based, baseline-based, and direction-dependent or direction-independent effects \citep{smirnov2011revisitingI}. %
For the scope of this paper only antenna-based calibration terms are considered, a simplification which is sensible for telescopes with a small field of view such as ALMA or the VLA. %
The crucial idea of this paper is to view the amplitude and phase corrections for each antenna as one-dimensional fields that are defined over time. %
These fields are discretized and regularized by a prior which states that the calibration solution for a given antenna is smooth over time. %
This removes the ambiguity of an interpolation scheme in between the calibrator observations and the subsequent application of self-calibration. %
Because \textsc{resolve} is an IFT algorithm, there is no notion of solution intervals, which are time bins in which traditional calibration algorithms bin the data \citep[see, e.g.,][]{kenyon2018cubical}. %
Instead IFT takes care of a consistent discretization of the principally continuous fields. %
Similarly, the sky brightness distribution is defined on a discretized two-dimensional space; only single-channel imaging is performed in this work. %

In practice, the current approach in the IFT community is to define a generative model that turns the degrees of freedom, which are learned by the algorithm into synthetic data that can be compared to the actual data in a squared-norm fashion (in the case of additive Gaussian noise). %
This approach is similar to the so-called radio interferometric measurement equation  (RIME; \citealt{hamaker1996understanding, perkins2015montblanc1, smirnov2011revisitingI}). %
Therefore, our notation closely follows the notation defined in \citet{smirnov2011revisitingI}. %
Calibration effects that are part of the RIME but left out for simplicity in this publication could in principle be integrated into the \textsc{resolve} framework. %

The \textsc{resolve} approach may be classified according to the notion of first, second, and third generation calibration established in \citet{noordam2010meqtrees}: it unifies cross-calibration (1GC), self-calibration (2GC), and imaging. Still it is to be strictly distinguished from existing approaches like \citet{kenyon2018cubical, cai2018uncertaintyI}, and \citet{salvini2014fast}. %
This is because it focuses on a strict Bayesian treatment combined with consistent discretization (one of the main benefits of IFT) and does not use computational speed as an argument to drop Bayesian rigidity. %

The actual posterior probability distribution of the joint imaging and calibration problem is highly non-Gaussian and therefore not easily storeable on a computer. %
In order to overcome this apparent problem the posterior is approximated by a multivariate Gaussian with full covariance matrix. %
The algorithm prescribes to minimize the Kullback-Leibler divergence (KL divergence) between the actual posterior and the approximate one which is the information gain between the two probability distributions. %
We use the variant of this known as Metric Gaussian Variational Inference (MGVI) \citep{knollmuller2019metric}. %

Together with this publication we contribute an implementation of \textsc{resolve} that is available under the terms of GPLv3.\footnote{\url{https://gitlab.mpcdf.mpg.de/ift/resolve}} %
It is based on the Python library NIFTy \citep{arras2019nifty5}, which is freely available as well. %

The paper is divided into four sections. Section~\ref{sec:algorithm} discusses the structure of likelihood and prior for the statistical problem at hand. %
This defines an algorithm which is verified on synthetic data in Section~\ref{sec:verification} and afterward applied to real data from the VLA in Section~\ref{sec:vla}. %
Section~\ref{sec:conclusions} finishes the paper with conclusions and a outlook for future work. %

\section{The algorithm}
\label{sec:algorithm}
\subsection{Bayes' theorem}
Every reconstruction algorithm needs a prescription of how the quantity of interest $s$ affects the data $d$. %
This prescription is called the data model. %
Combined with statistical information, this model defines the likelihood $\prob (d|s),$ which is a probability distribution on data realizations conditioned on a given realization of the signal $s$. %
Bayes' theorem,
\begin{align}
  \post  = \frac{\lh \prior}{\prob (d)},
\end{align}
requires us to supplement the likelihood with a prior probability distribution $\prob (s)$, which assigns a probability to each signal realization $s$. %
This distribution encodes the knowledge the scientist has prior to looking at the data. %
Since it is virtually impossible to visualize the posterior probability distribution $\prob (s|d)$ in the high dimensional setting of Bayesian image reconstruction we may compute the posterior mean and posterior variance as
\begin{align}
  m:=\langle s \rangle_{\post} :=&  \intds \post\, s,\\
  \langle |m-s|^2 \rangle_{\post} :=&  \intds \post\, |m-s|^2.
\end{align}
The notation $\intds$ is borrowed from statistical physics and means integrating over all possible configurations $s$. %
For a discussion on this measure in the continuum limit see \citet[section 1.8]{ensslin2018information}. %
In practice, this integral is discretized as follows: $\intds = \int\prod_i ds_i$ where $s_i$ refers to the pixel values of the discretized quantity $s$. %
The term $\prob (d)$ is independent of $s$ and serves as a normalization factor. %
It expresses the probability to obtain the data irrespective of what the signal is, i.e. $\prob (d) = \intds \prob(d,s)$.

In the following we describe the data model and implied likelihood employed by \textsc{resolve}. %
This includes the assumptions \textsc{resolve} makes about the measurement process. %
Afterward, \textsc{resolve}'s prior is discussed. %
For definiteness the notation established in \citet{smirnov2011revisitingI} is used.

\subsection{Data model and likelihood}
The measurement equation of a radio interferometer can be understood as a modified Fourier transform followed by an application of data-corrupting terms, the terms which need to be solved for in the calibration procedure. %
Assume that the data is only corrupted by so-called antenna-based direction-independent effects. %
Then \citet[equation 18]{smirnov2011revisitingI} is written as
\begin{align}
  V_{pq} =  G_p \left(\int  \frac{B(l,m)}{n(l,m)}\, e^{-2\pi i[u_{pq}l+v_{pq}m+w_{pq}(n(l,m)-1)]}\, dl\,dm \right) G_q^\dagger,
  \label{eq:fullrime}
\end{align}
where
\begin{itemize}
\item $l,m$: Direction cosines on the sky and $n(l,m)=\sqrt{1-l^2-m^2}$.
\item $p,q\in \{1, \ldots, N_a\}$: Antenna indices where
  $N_a$ is the total number of antennas of the interferometer.
\item $V_{pq} \in \mathbb C^{2\times 2}$: Visibility for antenna pair $(pq)$.
\item $(u_{pq}, v_{pq}, w_{pq})$: Vector that connects antenna $p$ with antenna $q$. %
  The coordinates $u_{pq}$ and $v_{pq}$ are aligned with $l$ and $m$, respectively. %
  The value $w_{pq}$ is perpendicular to both and points from the interferometer toward the center of the field of view. %
\item $G_p \in \mathbb C^{2\times 2}$: Antenna-based direction-independent calibration effect.
\item $B \in \mathbb R^{2\times 2}$: Intrinsic sky brightness matrix. Since only the Stokes~I component is considered in this publication, this matrix is proportional to the identity matrix. %
\end{itemize}
Equation \eqref{eq:fullrime} can be understood as a Fourier transform of the sky brightness distribution, which is distorted by the terms involving $n(l,m)$ and corrupted by the calibration terms $G_p$. %
For the purpose of this publication we make the following simplifying assumptions: %
First, only the total intensity $I$ is reconstructed. %
Second, $G_p$ is assumed to be diagonal, which states that there is no significant polarization leakage and especially no time-variable leakage. %
Finally, the temporal structure of the data is needed for the construction of the prior. %
Therefore, a time index is added to the above expression that is written as
\begin{align}
  V_{pqt} =  G_p(t)\left(  \int  \frac{B(l,m)}{n(l,m)} \, e^{-2\pi i[u_{pq}l+v_{pq}m+w_{pq}(n(l,m)-1)]}\, dl\,dm   \right) G_q^\dagger(t) ,
  \label{eq:rime}
\end{align}
where $G_p(t)$ are diagonal matrices and $B(l,m)$ is a diagonal matrix, which is proportional to unity in polarization space. %
We note that $G_p(t)$ needs to absorb the $V$-term from \eqref{eq:fullrime}, which is possible as long as polarization leakage is not too time variable. %
The $w$-term can be taken care of by $w$-stacking \citep{offringa2014wsclean}, which means that the range of possible values for $w_{pq}$ is binned linearly such that the integral becomes an ordinary Fourier transform.
Technically, the non-equidistant Fourier transform in \eqref{eq:rime} is carried out by the NFFT library \citep{keiner2009using} in our \textsc{resolve} implementation. %

All in all, \eqref{eq:rime} prescribes how to simulate data $V_{pqt}$ given calibration solutions $G_p(t)$ and an inherent sky brightness distribution $B(l,m)$, which is what we wanted. %
In order to declutter the notation in the following let us denote the quantities of interest by $s =\left(G_p(t), B(l,m)\right)$ and the map $R$ such that $V_{pqt}=R(s)$.

The commonly used data model is the following: $d=R(s)+n$. %
It assumes additive Gaussian noise \citep{thompson1986interferometry}. %
Let $N$ be a diagonal noise covariance matrix with the noise variances on its diagonal and $\gauss{s-m}{S}$ refers to a Gaussian random field with mean $m$ and covariance matrix $S$. %
Then, the additive noise can be marginalized over to arrive at an expression for the likelihood
\begin{align}
  \prob (d|s) &= \intdn \prob (d|s,n)\, \prob (n) \\
              &= \intdn \delta (n-(d-R(s))) \, \gauss{n}{N} \\
              &= \gauss{d-R(s)}{N}.
\end{align}
The likelihood distribution $\lh$ contains all information about the measurement device and the measurement process that the inference algorithm will take into account.

We conclude the discussion on data and likelihood with three remarks: First, the likelihood does not depend on the statistical method at hand. %
All simplifications being made are rooted in practical reasons in the implementation process. %
There is no fundamental reason for not taking, for instance, a more accurate noise model or a more sophisticated calibration structure into account.

Second, the employed notation already hints at the goal of describing an algorithm that jointly calibrates and images: %
the generalized response function $R$ takes at the same time the calibration parameters $G_p(t)$ and the intrinsic sky brightness distribution $B$ as an argument.

Finally, we consider what happens if the telescope alternates between observing the science target and observing a calibration source. %
Then, both the data set and the intrinsic sky brightness consists of two parts and the likelihood separates into
\begin{align}
  \label{eq:doublelikelihood}
  \lh = \prob (d_c|s)\, \prob (d_t|s)
\end{align}
From the likelihood perspective, calibration and science source are two separate things. %
However, as soon as the one-dimensional calibration fields are supplemented by a prior that imposes temporal smoothness the degrees of freedom regarding the science target and calibration target interact. %
This solves the problem of applying interpolated calibration solutions in traditional cross-calibration in a natural way.

\subsection{Prior}
Turning to the prior probability distribution, we note that the technical framework in which \textsc{resolve} is implemented allows for a variety of different priors, which may supersede that presented in this paper.

As stated before $G_p(t)$ are assumed to be diagonal,
\begin{align}
  G_p(t) = \begin{pmatrix} g^{(0)}_p(t)&0\\0&g^{(1)}_p(t) \end{pmatrix}.
\end{align}
The elements of this matrix are functions defined over time and take the following complex nonzero values:\footnote{$\mathbb C^*$ are the units of $\mathbb C$, i.e., $\mathbb C^* := \mathbb C \setminus \{0\}$.} %
\begin{align}
  g^{(i)}_p: [t_0,t_1]\to \mathbb C^*, \quad i \in \{0,1\},\, p\in\{1,\ldots,N_a\}.
\end{align}
The natural way of parameterizing a function taking values in $\mathbb C^*$ is in polar coordinates, i.e.,
\begin{align}
  g^{(i)}_p(t) = \exp \left( \lambda^{(i)}_p(t) + i \phi^{(i)}_p (t)\right),
\end{align}
where $\lambda^{(i)}_p: [t_0,t_1]\to \mathbb R$ and $\phi^{(i)}_p: [t_0,t_1]\to \nicefrac{\mathbb R}{2\pi \mathbb Z}$. %
The modulus and phase of the complex gains $g_p^{(i)}$ have different physical origins. %
The modulus describes a varying amplification of the signal in the antenna electronics, which is rooted amongst others in fluctuating temperatures of the receiver system. %
Varying phases stem from fluctuations in the atmosphere. %
Therefore, these two ingredients of $g_p$ have differing typical timescales a priori. %

The prior knowledge on $\lambda_p^{(i)}$ and $\phi_p^{(i)}$ is the following: $\{\lambda_p^{(i)}\}, \{\phi_p^{(i)}\}$, respectively, share a typical behavior over time for all antennas $p$, both of which are not known a priori and need to be inferred from the data as well. %
This typical behavior does not change over time. %
Additionally, all $\lambda_p^{(i)}, \phi_p^{(i)}$ evolve smoothly over time. %
Mathematically, this can be captured by Gaussian random fields,
\begin{align}
  \prob \left(\left.\left(\lambda_p^{(i)}, \phi_p^{(i)}\right)_{i,p}\right| \Lambda, \Phi \right) =\prod_{i, p} \gaussauto{\lambda_p^{(i)}}{\Lambda} \gaussauto{\phi_p^{(i)}}{\Phi},
\end{align}
where $\Lambda, \Phi$ is defined such that the Gaussian random fields obey homogeneous but still specifically unknown statistics. %
This means that not only the calibration solutions themselves but also their prior correlation structure is inferred. %
For this a prior on the covariances needs to be supplemented: $\prob (\Lambda), \prob(\Phi)$. %
In Section~\ref{sec:correlatedfields} we descibe how to set up the prior on $\Lambda$ and $\Phi$ such that they implement homogeneous statistics and which parameters they take.

Next, let us discuss the prior on the sky brightness distribution $B(l,m)$. %
We recall that the matrix $B(l,m)$ is assumed to be diagonal and proportional to unity, i.e.,
\begin{align}
  B(l,m) = \begin{pmatrix}b(l,m)&0\\0&b(l,m) \end{pmatrix},
\end{align}
where $b(l,m): [l_{\mathrm{min}}, l_{\mathrm{max}}]\times [m_{\mathrm{min}}, m_{\mathrm{max}}]\to \mathbb R_{>0}$ map the field of view to the set of positive real numbers since sky brightness is inherently a positive quantity.\footnote{We note the difference to Högbom's CLEAN, which has positivity not built in \citep{hogbom1974aperture}.} %
For the scope of this publication, the sky brightness contains only a diffuse component. %
It shall be modeled similarly to the modulus of the calibration terms: it is strictly positive a priori, smooth over its domain and may vary over large scales. %
Therefore, we define $b(l,m) = e^{\psi(l,m)}$ and let $\psi (l,m)$ be a two-dimensional Gaussian random field with correlation structure $\Psi,$ which is going to be inferred as well:
\begin{align}
  \prob (\psi|\Psi) = \gauss{\psi}{\Psi}.
\end{align}
All in all, the basic structure of the priors on all terms appearing in \eqref{eq:rime} has been described apart from the construction of the prior on all covariance matrices, which is the objective for the next section. %

\subsection{Correlated fields}
\label{sec:correlatedfields}

To account for correlations of a Gaussian distributed field $\psi$ the following statements are assumed to be true:
\begin{enumerate}
  \item The autocorrelation of $\psi$ can be characterized by a power spectrum $P_\Psi(|k|)$, where $k$ is the coordinate of the Fourier transformed field. \label{item:power-spectrum} %
  \item The power spectrum $P_\Psi(|k|)$ is a positive quantity that can vary over many orders of magnitudes. \label{item:power-positive} %
  \item Physical power spectra are falling with $|k|$, typically according to a power law. \label{item:power-law-falling} %
  \item Given enough data, it is possible to infer any kind of differentiable power spectrum. \label{item:representative-power} %
\end{enumerate}
Note that the first assumption is equivalent to the seemingly weaker assumptions: %
\begin{itemize}
\item In absence of data, there is no special direction in space or time, i.e., a priori the correlation of the field is invariant under rotations. %
\item In absence of data, there is no special point in space or time, i.e., a priori the correlation of field values is invariant under shifts in space or time. %
\end{itemize}
The fact that homogeneous and isotropic correlation matrices are diagonal in Fourier space and can be fully characterized by a power spectrum is known as the Wiener-Khinchin theorem \citep{Wiener, Khintchin}. %

It is assumed that $\psi$ as well as its power spectrum $P_\Psi(|k|)$ are unknown. %
Therefore, both need a prior that may be formulated as generative model: an operator that generates samples for $\psi$ and its square root power spectrum (henceforth called amplitude spectrum) from one or multiple white Gaussian fields. %
Formulating a prior as a generative model has several theoretical and practical advantages \citep{knollmueller2018encoding}. %

\begin{figure}
  \centering
  \includegraphics{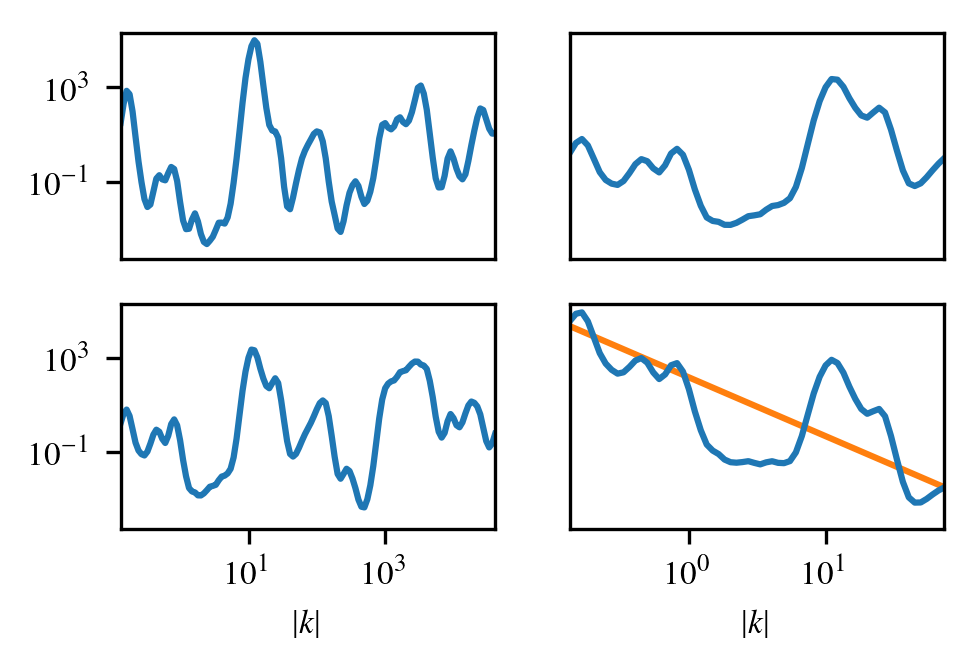}
  \caption{Steps of the generative process defined in Eq.\ \eqref{eq:amplitude-operator}. %
  Top left: Smooth, periodic field defined on the interval $[t_0,2t_1-t_0]$. %
  Bottom left: (anti-)symmetrized version of the above. %
  Top right: Projection of the symmetrized field to half of the original domain $[t_0,t_1]$. %
  Bottom right: Resulting double logarithmic amplitude spectrum after addition of the power law (orange) to the above.}
  \label{fig:correlatedfields}
\end{figure}

We propose the following ansatz for an operator that converts independent normal distributed fields, $\phi$ and $\tau$ to the amplitude spectrum of the correlated field $\psi$. %
This operator is called amplitude operator $A_C$ (see Figure~\ref{fig:correlatedfields} for an illustrative example), i.e. %
\begin{multline}
A_C(\tau, \phi) = (\text{Exp}^* \text{Exp})\bigg(0.5\cdot\Big[\log(k)(\sigma_m\tau_m +\bar{m})+ \sigma_{y_0}\tau_{y_0}+\bar{y}_0 \\
  +(\text{sym}\circ\widetilde{\mathcal F}_{\log(k)t})(\text{cp}(t)\cdot\phi(t))\Big]\bigg),\label{eq:amplitude-operator} %
\end{multline}
where $C = ( a, t_0, \bar m, \bar y_0, \sigma_m, \sigma_{y_0}, \alpha, \beta)$ denotes the tuple of parameters (all real numbers), $\text{Exp}^*$ denotes the pullback of a field by the exponential function acting on $\log(|k|)$\footnote{Let $\phi: U\to V$ with $U, V \subseteq \mathbb R$ open and $f: V\to\mathbb R$ a smooth function, i.e., a field. Then $(\phi^*f)(t):= f(\phi(t))$ denotes the pullback of $f$ by $\phi$. In other words, the field $f$ is transformed to a different coordinate system whose coordinates are related to the original one by $\phi$.}, $\text{Exp}$ denotes exponentiation of the field values, $\widetilde{\mathcal F}_{\log(k)t}$ denotes the Fourier transform of a space with coordinates $t$ to the logarithmic coordinates $\log(k)$ of the power spectrum, $\bar{m}$ and $\bar{y}_0$ are the slope and the $y$-intercept of the a priori mean power law, $\text{sym}$ is an (anti-)symmetrizing operation defined to operate on a field $\phi$ over the interval $[t_0,t_1]$ as
\begin{align}
  2\cdot \text{sym}(\phi)(x) = \phi(x)-\phi(2t_1-t_0-x), %
\end{align}
for  $x \in (t_0,2t_1-t_0)$. %
In words, $\text{sym}$ mirrors the field and subtracts it from itself, then restricts the domain to half the original size. %
Finally, $\text{cp}$ is the log-cepstrum,
\begin{align}
  \text{cp}(t) = a\cdot\left(1+\left(\nicefrac{t}{t_0}\right)^{-2}\right) .
\end{align}

Let us show that \eqref{eq:amplitude-operator} meets the requirements stated at the beginning of section \ref{sec:correlatedfields}. %
Requirement~\ref{item:power-spectrum} is trivial. %
Requirement~\ref{item:power-positive} is met since the amplitude spectrum is constructed by applying an exponential function to a Gaussian field. %
Thus, all values are positive and can vary over several order of magnitudes. %

To requirement~\ref{item:power-law-falling}: %
In absence of data, the mean of the inferred white fields $\phi$ and $\tau$, to which the amplitude operator is applied, remains zero. %
For $\phi=0$ and $\tau=0$, \eqref{eq:amplitude-operator} becomes %
\begin{align}
  (\text{Exp}^*\text{Exp})(0.5\cdot [\bar{m}\log(k)+ \bar{y}_0 ]),
\end{align}
which is the equation for a power law with spectral index $\bar{m}$. %
A preference for falling spectra can be encoded by choosing the hyperparameter $\bar{m}$ to be negative. %

To requirement~\ref{item:representative-power}: %
Let us show that any differentiable function lies in the image space of the amplitude operator. %
This implies that any differentiable amplitude spectrum can be inferred given enough data. %
Let $\phi$ be an arbitrary smooth field over the interval $[t_0,t_1]$ and $\phi_{\text{sym}}$ be a smooth field that has a point symmetry at $(t_1, \phi(t_1))$ and is defined on the interval $[t_0,2t_1-t_0]$ as
\begin{align}
  \phi_\text{sym}(t) =
  \begin{cases}
    \phi(t) &\text{for } t\in [t_0,t_1], \\
    2\phi(t_1) - \phi(2t_1-t) &\text{for } t\in (t_1,2t_1-t_0].
  \end{cases}
\end{align}
The function $\phi_{\text{sym}}$ is a continuous and differentiable continuation of $\phi$ at $t_1$. %
Now, we decompose $\phi_\text{sym}$ into a linear part and a residual term: %
\begin{align}
  \phi_\text{sym}(t) &= m\cdot(t-t_0) + y_0 + \phi_\text{res}(t),
\end{align}
where
\begin{align}
 y_0 &= \phi_\text{sym}(t_0), \\
  m &= \frac{\phi_\text{sym}(2t_1-t_0) - \phi_\text{sym}(t_0)}{2(t_1-t_0)},\\
   \phi_\text{res}(t) &= -m\cdot(t-t_0) - y_0 + \phi_\text{sym}(t).
\end{align}
The residual term is a differentiable periodic function, i.e.,
\begin{align}
  \begin{split}
  \phi_\text{res}(t_0) &= \phi_\text{res}(2t_1+t_0)\\
  \Leftrightarrow\phi_\text{sym}(t_0) - \phi_\text{sym}(t_0) &= -\phi_\text{sym}(2t_1-t_0) + \phi_\text{sym}(t_0) \\&\quad- \phi_\text{sym}(t_0) + \phi_\text{sym}(2t_1+t_0)
  \end{split}\\
  \begin{split}
  \phi_\text{res}^\prime(t_0) &=
  \phi_\text{res}^\prime(2t_1+t_0)\\
  \Leftrightarrow\phi^\prime(t_0) - m &=
  \phi_\text{sym}^\prime(2t_1+t_0)-m\\
  \Leftrightarrow\phi^\prime(t_0) - m &=
  \phi^\prime(t_0) - m.
  \end{split}
\end{align}
Thus, $\phi_\text{res}$ can be represented in Fourier space by a field that falls of at least with second order. %
This is exactly how $\phi_\text{res}$ is represented in \eqref{eq:amplitude-operator}. %
Assuming that the mean and the slope of the linear part are well represented by its prior distribution, it is indeed possible to represent any kind of differentiable amplitude spectrum. %
All in all, all four requirements are met by \eqref{eq:amplitude-operator}. %

There remains one unconstrained degree of freedom, the value of the power spectrum at $|k| = 0$, the zero mode. %
As the zero mode describes the magnitude of the overall logarithmic flux, it is decoupled from the remaining spectrum and should have its own prior. %
This value is fixed by imposing an inverse gamma prior on the zero mode, which restricts it to be a positive quantity, while still allowing for large deviations. %

To sum up, the amplitude operator depends on the following eight hyper parameters: %
\begin{itemize}
\item $a$, $t_0$: The amplitude parameter and cutoff scale of the log-cepstrum.
\item  $\bar{m}$, $\bar{y}_0$: The prior means for the slope and the height of the power law.
\item  $\sigma_m$, $\sigma_{y_0}$: The corresponding standard deviations.
\item  $\alpha$, $\beta$: The shape and scale parameter of the inverse gamma prior for the zero mode.
\end{itemize}

We note that the assumptions made at the beginning of section \ref{sec:correlatedfields} apply to a wide variety of processes, regardless of their dimensionality. %
This generic correlated field model has already been successfully used in a number of synthetic and real applications \citep{leike2019charting, knollmueller2018encoding, knollmuller2019metric, knollmuller2018separating, hutschenreuter2019galactic}. %
In \textsc{resolve}, the amplitude operator is used as a prior for the amplitude spectra of the antenna calibration fields and the image itself. %

\subsection{ Full algorithm}
In the aforegoing sections, the full likelihood and prior are described. %
Now, we stack all the ingredients together to build the full algorithm. %
Let us assume that the data set consists out of two alternating observations: observations of a calibrator source and observations of the science target. %
This means that the likelihood splits into two parts as indicated in \eqref{eq:doublelikelihood}. %
In contrast to the sky brightness distribution of the science target that of the calibrator $B_c$ is known: it is a point source in the middle of the field of view. %
The sky brightness distribution of the science target is reconstructed.

The full likelihood takes the form
\begin{align}
  \prob (d_t|\xi)\,\prob (d_c|\xi) &=\prod_{a\in \{t,c\}} \gauss{d_a-R_a(\{G_p^{(i)}\}, B_a)}{N_a\otimes \mathds 1},\label{eq:withtensor}\\
  B_t &= \exp \circ \mathcal F \circ (\xi_B \cdot A^B),\\
  G_p^{(i)}&= \begin{pmatrix} g_p^{(i)}&0\\0&g_p^{(i)} \end{pmatrix},\\
  g_p^{(j)} &= \exp \left(\lambda_p^{(j)}+i\phi_p^{(j)}\right),\\
  \lambda_p^{(i)} &= Z \circ \mathcal F \circ (\xi_{\lambda_p^{(i)}}\cdot A^\lambda),\\
  \phi_p^{(i)} &= Z \circ \mathcal F \circ (\xi_{\phi_p^{(i)}}\cdot A^\phi),\\
  A^B &= A_{C_B}(\xi_{A_B}),\\
  A^\lambda &= A_{C_\lambda}(\xi_{A_\lambda}),\\
  A^\phi &= A_{C_\phi}(\xi_{A_\phi}),
\end{align}
where $C_x$ denote the tuple of parameters of the respective amplitude operator, $Z$ is a padding operator. %
The unit matrices in \eqref{eq:withtensor} is a $2\times 2$ matrix acting on the same space as the sky brightness matrix $B$. %
The tuple of all excitation fields is called $\xi$, where
\begin{align}
  \xi &= \left(    \xi_B ,\xi_{A_B}, \xi_{A_\lambda}, \xi_{A_\phi},  \xi_{\lambda_0^{(0)}},\ldots, \xi_{\lambda_{N_a}^{(1)}}, \xi_{\phi_0^{(0)}},\ldots, \xi_{\phi_{N_a}^{(1)}} \right) .
\end{align}
As discussed before this model is set up such that the excitation fields $\xi$ have white Gaussian statistics a priori,
\begin{align}
  \prob (\xi) = \gauss{\xi}{\mathds 1}.
\end{align}
The posterior probability distribution is given by
\begin{align}
  \label{eq:posterior}
   \prob (\xi| d_t, d_c) \propto \prob (d_t,d_c,\xi) = \prob (d_t|\xi)\,\prob (d_c|\xi) \, \prob (\xi).
\end{align}
Finally, the statistical model that is employed in this publication is fully defined. %

\subsection{Inference algorithm}
The probability distribution \eqref{eq:posterior} has too many degrees of freedom to be represented on a computer. %
The \textsc{resolve} algorithm solves this problem by approximating this full posterior distribution by a multivariate Gaussian distribution whose covariance is equated with the inverse Fisher information metric. %
The latter can be represented symbolically alleviating the need for an explicit storage and handling of otherwise prohibitively large matrices. %
This algorithm is called MGVI and is described in full length in \citet{knollmuller2019metric} and implemented in NIFTy.\footnote{\url{https://gitlab.mpcdf.mpg.de/ift/nifty}} %
The following is an inexhaustive outline of \citet{knollmuller2019metric}. %

The algorithm MGVI prescribes to minimize the KL divergence\footnote{Also known as discrimination information.} between the actual posterior and approximate posterior such that %
\begin{align}
  \mathrm{KL}(\mathcal P_1 || \mathcal P_2) = \int\mathcal Ds\, \mathcal P_1 (s) \log \left( \frac{\mathcal P_1 (s)}{\mathcal P_2(s)}\right),
\end{align}
where $\mathcal P_1$ is more informed compared to $\mathcal P_2$. %
However, it is apparent that it is virtually impossible to perform the integration with respect to the posterior distribution as integration measure. %
Therefore, MGVI exchanges the order of the arguments of the KL divergence such that the integral can be approximated by samples of the approximate posterior, i.e., %
\begin{align}
  F[\xi] = \left\langle \mathcal H(\xi+x,d) \right\rangle_{x\curvearrowleft\gauss{x}{D(\xi)}},
\end{align}
where $\mathcal H(\xi, d) := -\log \prob (\xi,d)$ is the information Hamiltonian and $D(\xi)$ the Fisher information. The parameter %
$F[\xi]$ is a cost function that can be minimized with respect to $\xi$. %
Suitable (2nd order) minimizers are provided by NIFTy. %

With the help of the above approximation scheme we gets a computational handle on the posterior. %
The drawbacks of this approach include the uncertainty estimate of MGVI sets a lower bound on the variance of the posterior and it is not suited for extremely non-Gaussian and especially multimodal probability distributions. %
  But we note that the posterior is approximated with a Gaussian in the space on which the parameters are defined. %
  After processing the parameters through nonlinearities as discussed in this section the actual quantities of interest such as the sky brightness distribution are not Gaussian distributed any more and may even have multiple modes. %
   A detailed discussion on the abilities of MGVI is provided in \citet{knollmuller2019metric}. %

\section{Verification on synthetic data}
\label{sec:verification}
This section is devoted to the verification of the algorithm, i.e., the reconstruction of a synthetic sky brightness distribution from a simulated observation and artificial noise. %
The setup is described followed by a comparison of the ground truth and the reconstruction. %
Application to real data, where effects that are not modeled may occur and the ground truth is unknown, is presented in Section~\ref{sec:vla}. %

\begin{figure}
  \centering
  \includegraphics{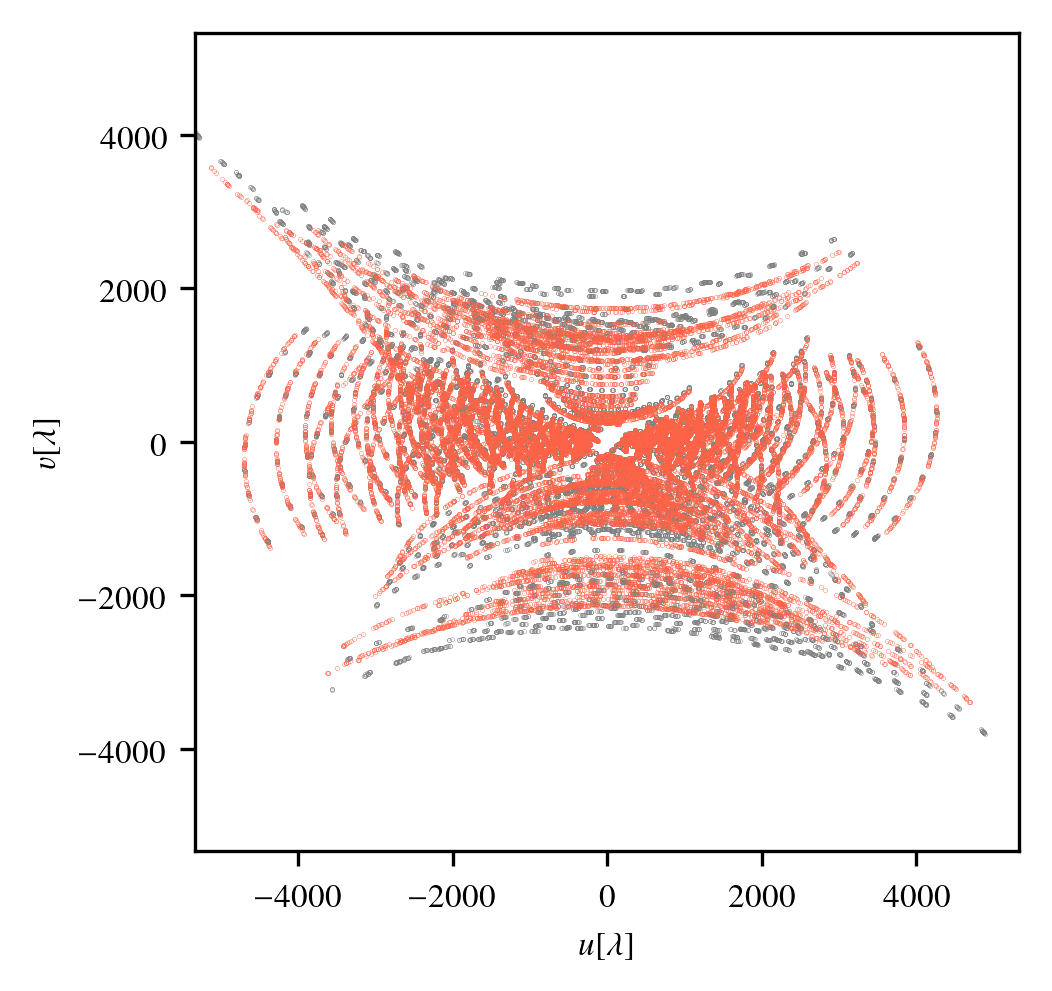}
  \caption{Random sample (30000 points) of uv coverage of a G327.6+14.6 (SN1006) observation with the VLA. The gray and red points indicate the uv coverage of the calibration source and science target, respectively.}
  \label{fig:uvcoverage}
\end{figure}

\begin{figure}
  \centering
  \includegraphics{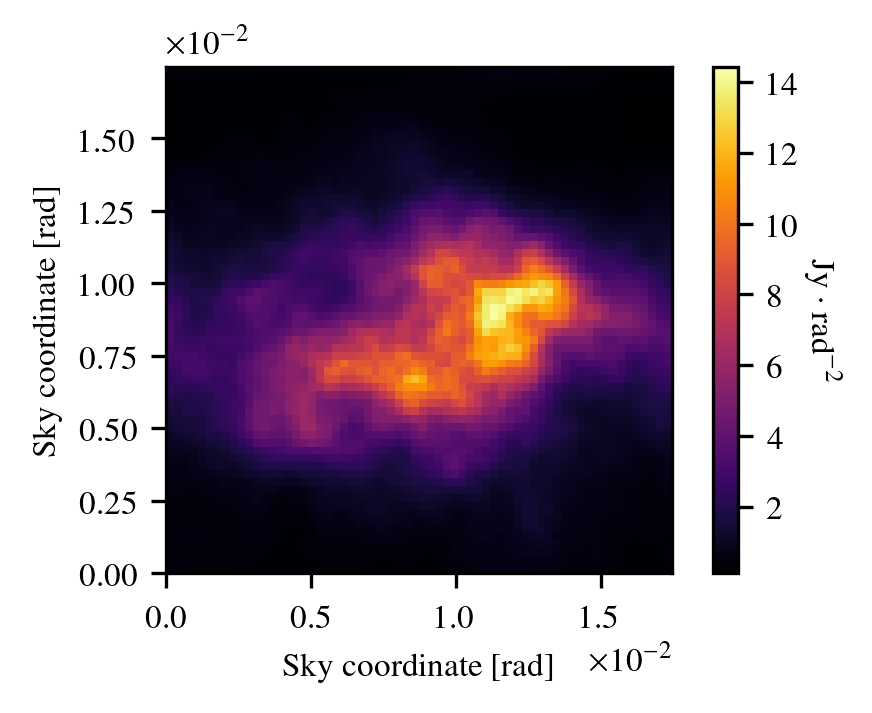}
  \caption{Synthetic observation: Ground truth sky brightness distribution $b(l,m)$ with 60' field of view.}
  \label{fig:mocksky_groundtruth}
\end{figure}

\begin{figure}
  \centering
  \includegraphics{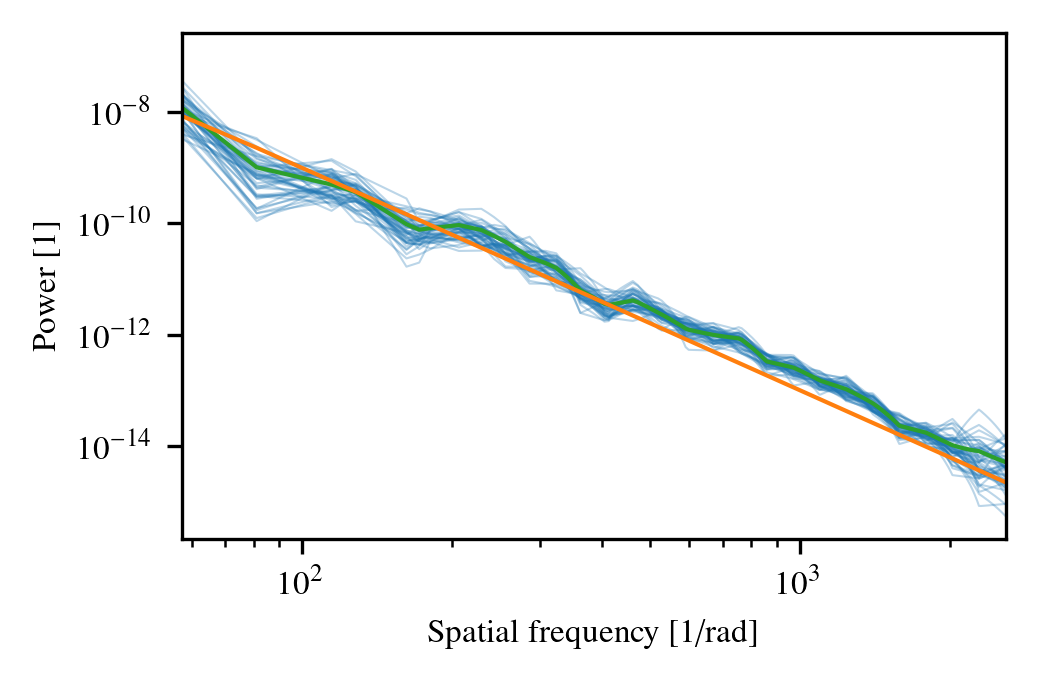}
  \caption{Synthetic observation: Power spectrum of log-sky brightness distribution. Orange: Ground truth; green: posterior mean; and blue: posterior samples.}
  \label{fig:mockpspec}
\end{figure}

\begin{figure}
  \centering
  \includegraphics{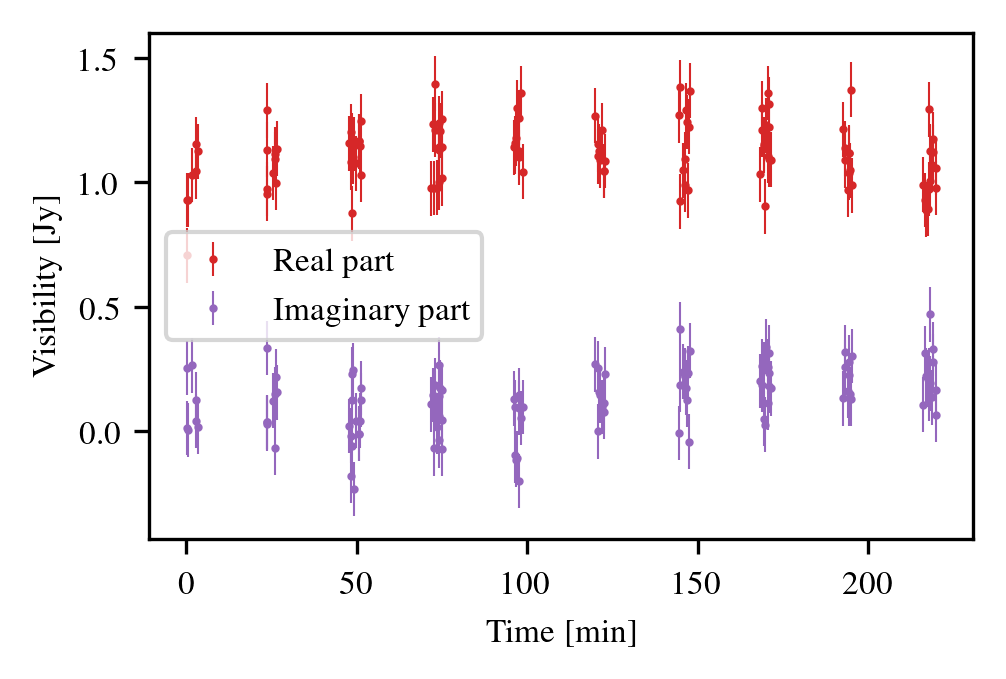}
  \caption{Synthetic observation: Visibilities of calibrator observation (polarization L, only visibilities of antennas 1 and 3). Thus, a constant value of $1+0j$~Jy is expected. All deviations from this are either noise or calibration errors. The error bars show the standard deviation on the data points.}
  \label{fig:mockdata}
\end{figure}

\begin{figure}
  \centering
  \includegraphics{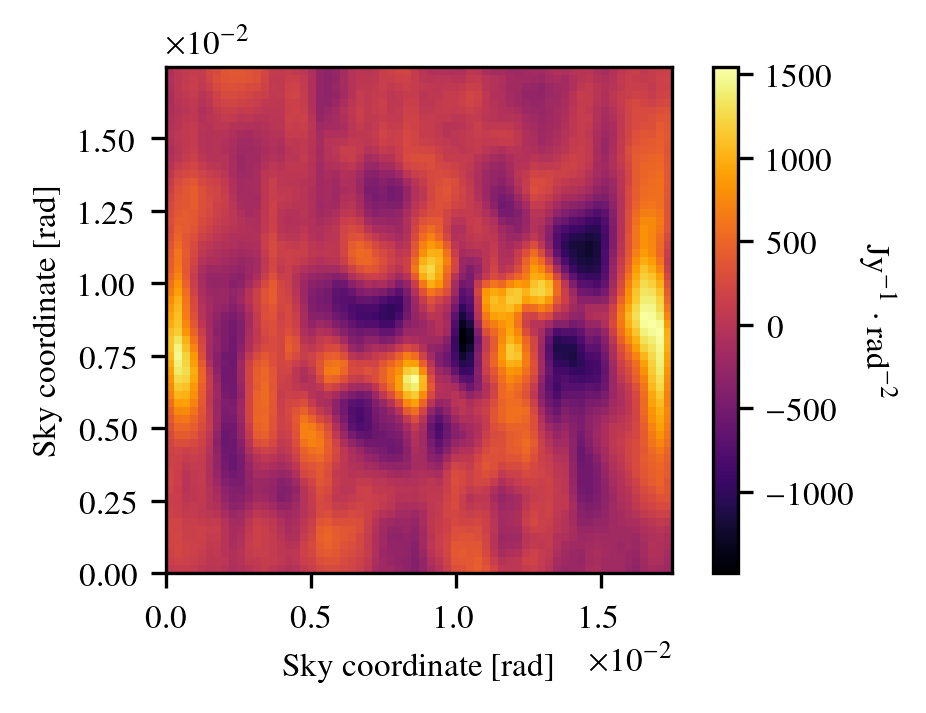}
  \caption{Synthetic observation: Information source $j=R_t^\dagger N_t^{-1} d_t$.}
  \label{fig:infosource}
\end{figure}

We employ a realistic uv coverage. %
It is an L-band observation of the source G327.6+14.6, also known as SN1006\footnote{ %
  VLA archive project code: AM0754, Jan 24, 2003, L-Band $1369.95$ MHz, CnD configuration. %
}.
For the purpose of this paper we randomly select 30000 visibilities from this data set to demonstrate that joint calibration and imaging is possible even without much data. (see Figure~\ref{fig:uvcoverage}). %
We use the field shown in Figure~\ref{fig:mocksky_groundtruth} as the synthetic sky brightness distribution. %
It is a random sample assuming the power spectrum shown in orange in Figure~\ref{fig:mockpspec}. %
The noiseless simulated visibilities are corrupted by noise whose level is visualized in Figure~\ref{fig:mockdata}. %
The resulting information source, i.e., the naturally weighted dirty image, is shown in Figure~\ref{fig:infosource}. %

\begin{figure*}
  \centering
  \includegraphics{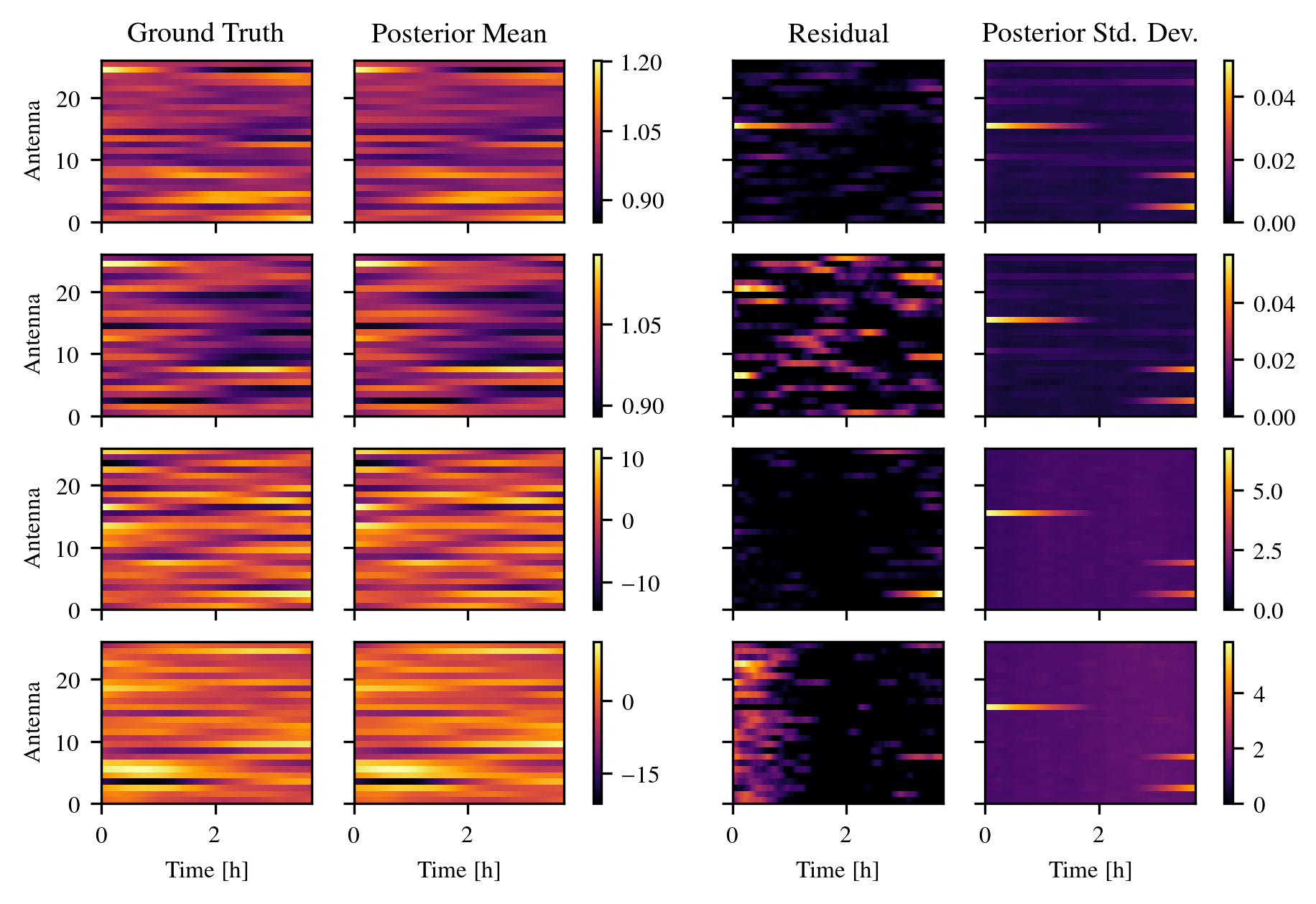}
  \caption{Synthetic observation: Calibration solutions. The first two rows show the amplitude and the bottom two rows show the phase calibration solutions. The first and the third row refer to LL-polarization and the second and last row to RR-polarization. The third column shows the absolute value of the difference between posterior mean and ground truth. The fourth column display the point-wise posterior standard deviation as provided by \textsc{resolve}. Amplitudes do not have a unit as they are a simple factor applied to the data. Phases are shown in degrees.}
  \label{fig:compcalib}
\end{figure*}

This synthetic observation is set up in a fashion such that the calibration artifacts are stronger and the noise level is higher as compared to real data (see~Section~\ref{sec:vla}) to demonstrate the capability of the \textsc{resolve} in bad data situations. %
The calibration artifacts that have been applied are visualized in Figure~\ref{fig:compcalib}. %

The \textsc{resolve} algorithm is run on this synthetic data to compare its output and uncertainty estimation to the (known) ground truth. %
The prior parameters are listed in Table~\ref{tab:synthetic}. %
Additionally, we choose a resolution of $64^2$ pixels for the sky brightness distribution with a field of view of $60'$ and 256 pixels for the calibration fields that are defined on a temporal domain. %
As the total length of the observation was $\sim 220$~min one temporal pixel is $\sim 50$~s long. %
These temporal pixels should not be confused with solution intervals of traditional calibration schemes where the data is binned on a grid and then the calibration parameters are solved for. %
In IFT fields are by their nature continuous quantities that are discretized on an arbitrary grid. %
For convenience a regular grid was chosen. %
Then the data provides information on each pixel that is propagated to the neighboring pixels through the prior; the calibration fields are assumed to be smooth over time. %
Therefore, the user is free to choose the resolution of the fields in IFT algorithms as long as it is finer than the finest structure that shall be reconstructed. %

\begin{table}
  \centering
  \begin{tabular}{ccccccccc}
    &$a$ & $t_0$ & $\bar m$ & $\bar y$ & $\sigma_m$ & $\sigma_{y_0}$ & $\alpha$ & $\beta$ \\
    \toprule
    $A$ & 2 & 2 & $-4$ & 5 & 1 & 3 & 4 & $5\cdot 10^{-3}$\\
    $\lambda$ & 1.5 & 1 & $-4$ & $-37$ & $0.5$ & 1 & 2 & 20\\
    $\phi$ & 1.5 & 1 & $-4$ & $-36$ & $0.5$ & 1 & 2 & 20\\
    \bottomrule
  \end{tabular}
  \caption{Synthetic observation: Prior parameters.}
  \label{tab:synthetic}
\end{table}

As pointed out \textsc{resolve} is a Bayesian algorithm whose output is not a single image of the observed patch of the sky but rather a probability distribution of all possible sky configurations. %
The MGVI algorithm approximates this non-Gaussian probability distribution with a Gaussian in the space of $\xi$, i.e., the eigenspace of the prior covariance. %
This again implies non-Gaussian statistics on quantities such as $b(l,m)$, $\lambda_p^{(i)}$, and $\phi_p^{(i)}$ since they depend in a nonlinear fashion on $\xi$. %
The only useful way to visualize this probability distribution is to analyze a finite number of samples from it which \textsc{resolve} can generate. %
A given set of samples can then be analyzed with standard statistical means such as the pixel-wise mean and variance. %

\begin{figure}
  \centering
  \includegraphics{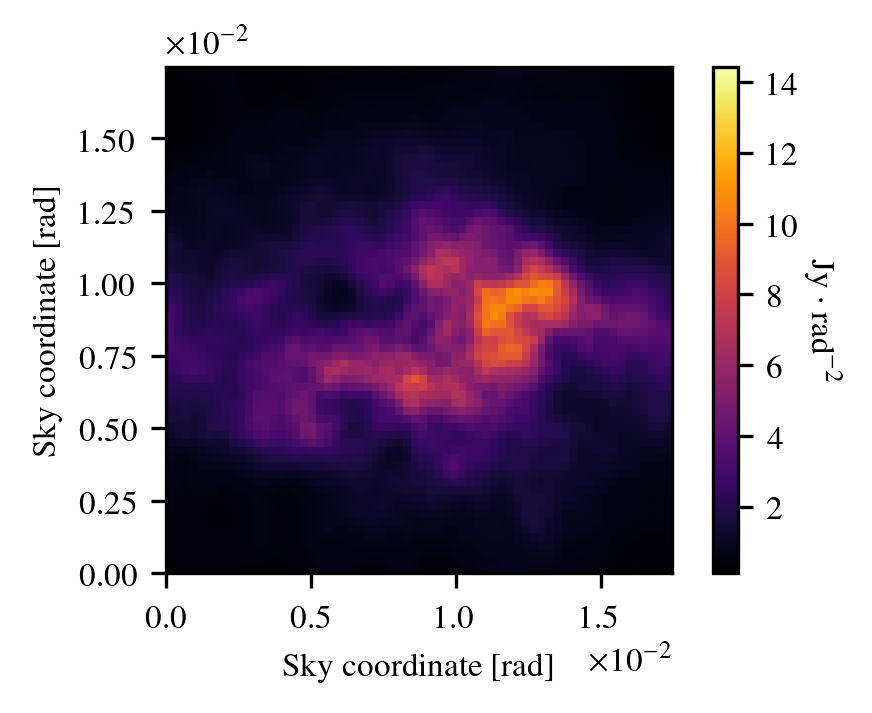}
  \caption{Synthetic observation: Posterior mean of sky brightness distribution $b(l,m)$.}
  \label{fig:mockpostmean}
\end{figure}

\begin{figure}
  \centering
  \includegraphics{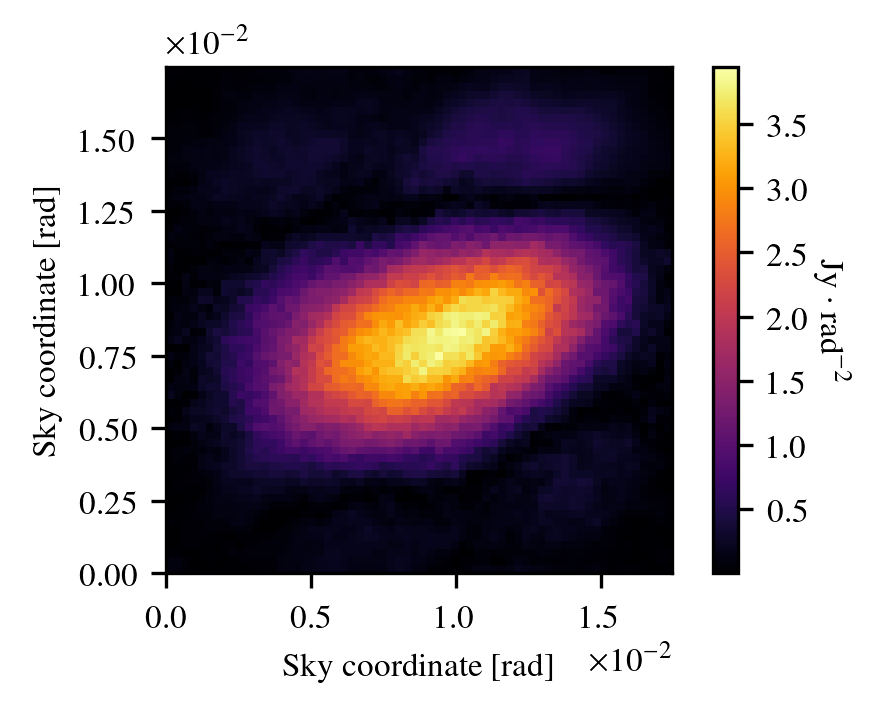}
  \caption{Synthetic observation: Absolute value of the difference between ground truth and posterior mean.}
  \label{fig:mockpostres}
\end{figure}

\begin{figure}
  \centering
  \includegraphics{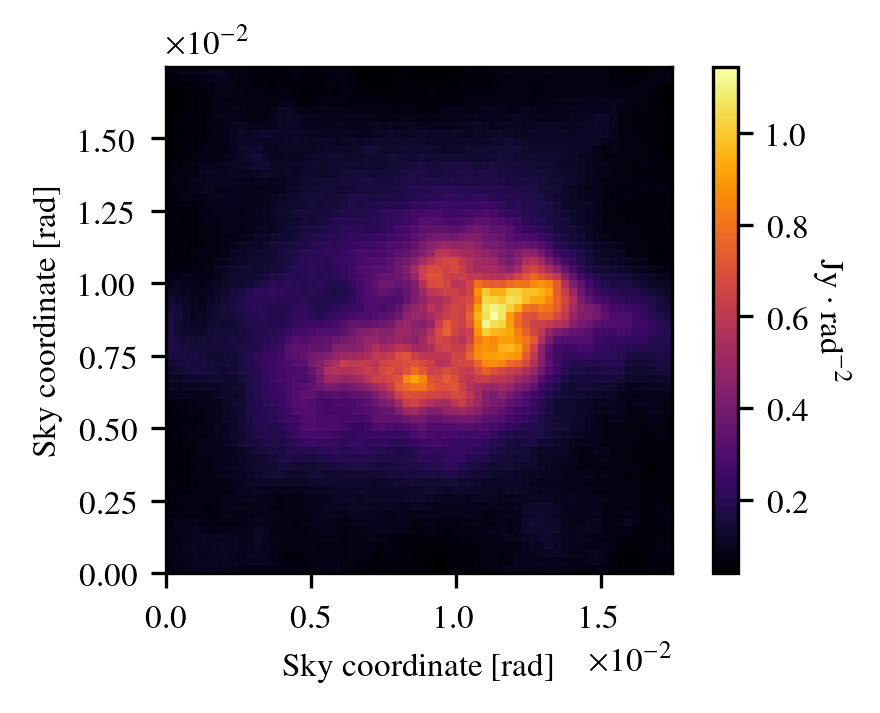}
  \caption{Synthetic observation: Pixel-wise standard deviation of posterior samples.}
  \label{fig:mockpostsd}
\end{figure}

Figures~\ref{fig:mockpostmean}, \ref{fig:mockpostres}, and \ref{fig:mockpostsd} show the posterior mean, the absolute value of the residual, the standard deviation of the sky brightness distribution, and a histogram of the residual divided by the standard deviation computed from 100 posterior samples, respectively. %
The algorithm has managed to perform the calibration correctly and to reconstruct the sky brightness distribution. %
The total flux of the ground truth (Fig.~\ref{fig:mocksky_groundtruth}) could not totally be recovered because of the noise on the synthetic measurement. %
Remarkably, the proposed uncertainty is a bit too small compared to the residuals which is what is to be expected from MGVI. %

Since \textsc{resolve} does not assume a specific power spectrum as prior for the reconstruction but rather learns it together with the sky brightness from the data, \textsc{resolve} also provides the user uncertainty on the power spectrum; see Figure~\ref{fig:mockpspec}. %
We note that the posterior variance on the power spectrum increases toward the boundaries of the plot. %
This is because interferometers do not provide information on scales larger than those that belong to the shortest baseline. %
On small scales an interferometer is limited by the noise level, which leads to an increased variance in the power spectrum on the right-hand side of the plot. %

Next, we turn to the calibration solutions. %
Figure~\ref{fig:compcalib} shows a comparison of the ground truth and the posterior provided by \textsc{resolve}. %
Since two polarizations are considered (LL and RR) for both the amplitude and the phase of the antenna-based calibration term, Figure~\ref{fig:compcalib} has four rows. %
On first sight, the posterior mean and the ground truth are indistinguishable by eye and the residuals and posterior standard deviation fit together nicely. %
There is a significant increase of the uncertainty for, for example, antenna 2 toward the end of the observation. %
This is because a flagged data set was used and that simply all data points involving this antenna have been flagged from the beginning of the observation up to $\sim 2$h. %

\begin{figure}
  \centering
  \includegraphics{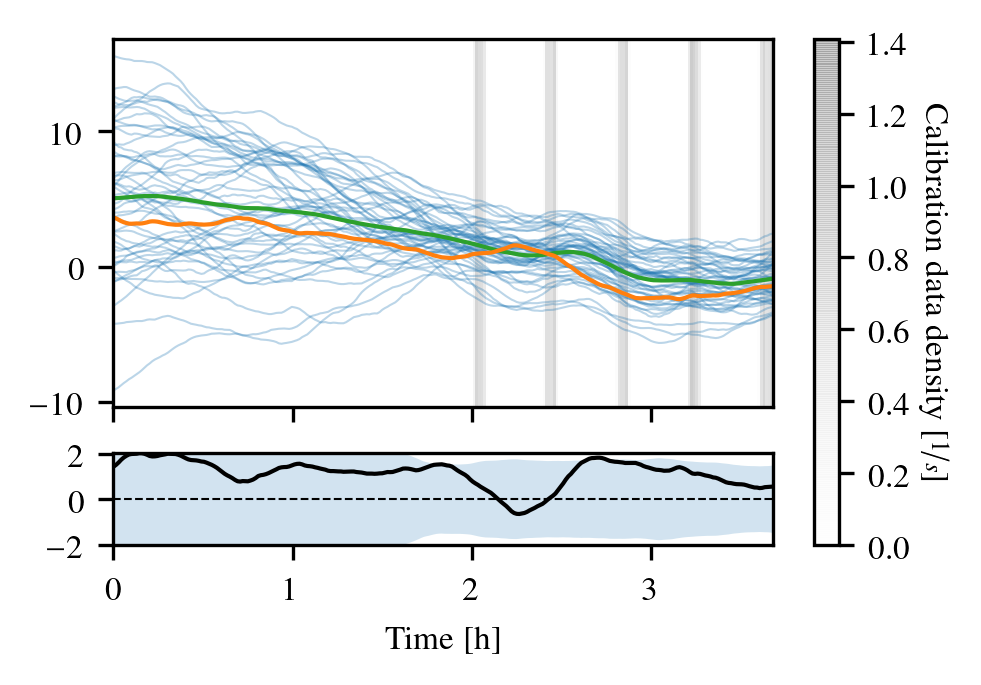}
  \caption{Synthetic observation: phase solutions for antenna 15 and polarization R. Colors as in Figure~\ref{fig:exampl} and all phases are plotted in degrees. On the left-hand side all data points have been flagged.}
  \label{fig:exph}
\end{figure}

\begin{figure}
  \centering
  \includegraphics{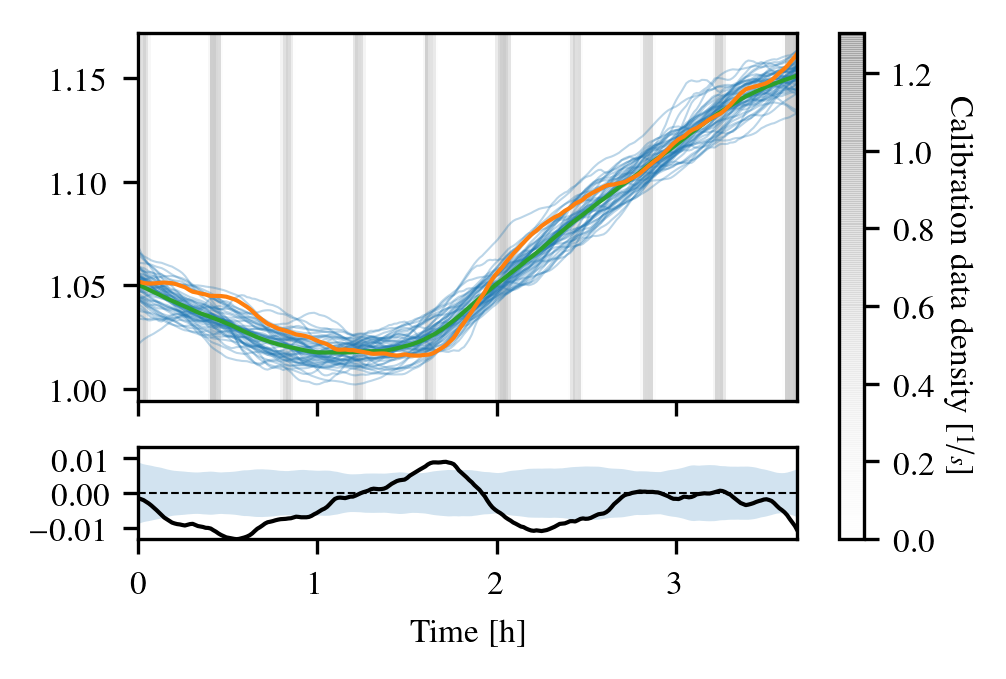}
  \caption{Synthetic observation: Amplitude solutions for antenna 0 and polarization L. Orange: Ground truth; green: sampled posterior mean; and blue: posterior samples. The calibration data density shows how many data points of the calibrator observation are available. We note that a Bayesian algorithm can naturally deal with incomplete data or data from different sources. The bottom plot shows the residual along with the pixel-wise posterior standard deviation.}
  \label{fig:exampl}
\end{figure}

To illustrate this more explicitly, Figures~\ref{fig:exph} and \ref{fig:exampl} show the calibration solution for one antenna, respectively. %
The ground truth lies within the bounds of uncertainty indicated by the samples. %
We note that all data points have been flagged on the left-hand side of Figure~\ref{fig:exph}. %
Since no information about the phase is available the only constraint is the prior, which enforces temporal smoothness. %
Consistently, the uncertainty increases where no information is available. %

\begin{figure}
  \centering
  \includegraphics{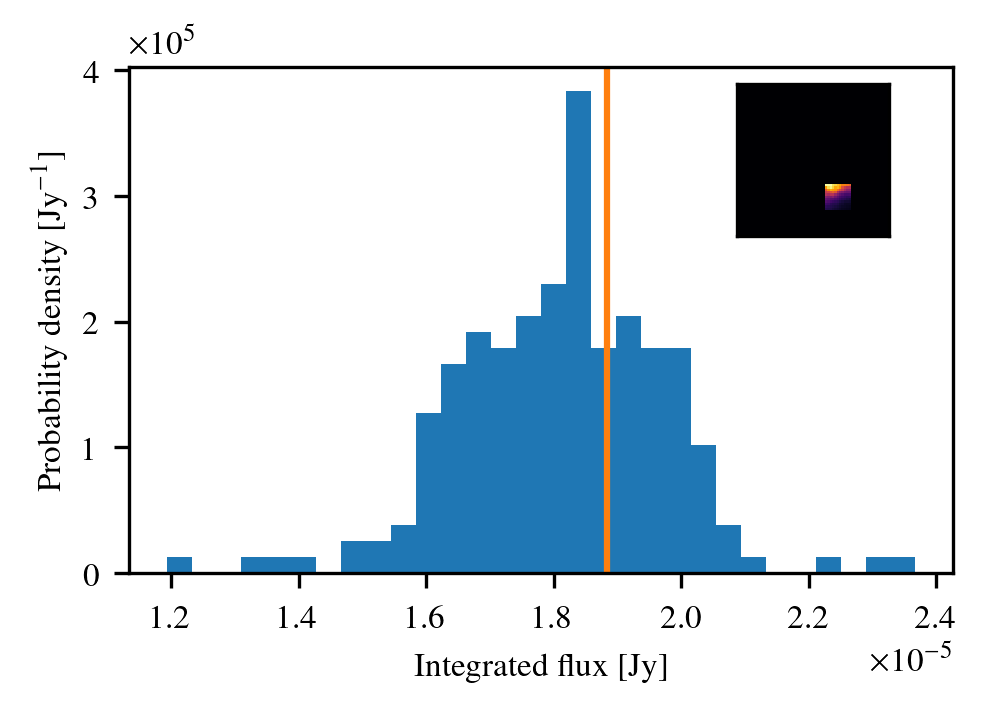}
  \caption{Synthetic observation: Histogram over samples of integrated flux in the region shown in the top right corner. Orange: Ground truth.}
  \label{fig:mockhisto}
\end{figure}

Finally, we demonstrate what kind of other information posterior samples can reveal. %
Say, a scientist is interested in the integrated flux over a certain region. %
In addition to the image, this integrated flux comes with an uncertainty that can be calculated by averaging over posterior samples of the sky brightness distribution. %
An example is shown in Figure~\ref{fig:mockhisto}. %
The scatter of the histogram is caused by the noise influence on the data, the (un)certainty of the calibration solutions, and ultimately the uv coverage. %
We are not aware of any other radio aperture synthesis algorithm that is able to provide this kind of probabilistic posterior information. %
All in all, the proposed method is able to recover the ground truth and is able to supplement it with an appropriate uncertainty estimation. %

\section{Application to VLA data}
\label{sec:vla}
We continue with an application of \textsc{resolve} to real data. %
To this end, take the VLA data set whose uv coverage and time stamps have already been used in the preceding section. %
Also, the resolution of all spaces is taken to be the same. %

\begin{figure}
  \centering
  \includegraphics{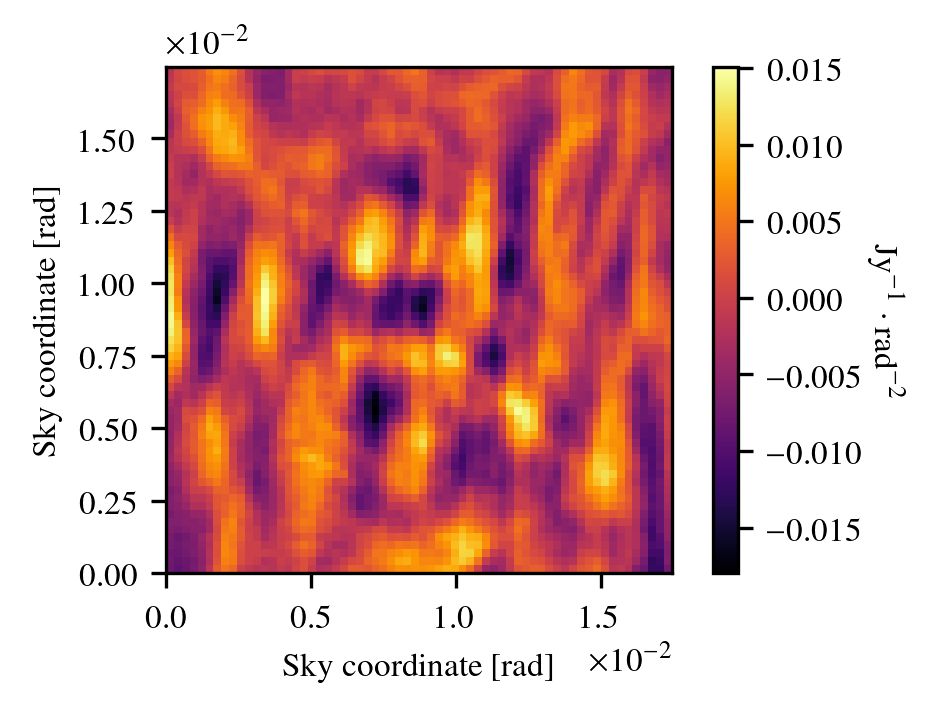}
  \caption{SN1006: Information source $j=R^\dagger N^{-1} d$.}
  \label{fig:snj}
\end{figure}

Starting from raw data, the first thing to look at is the information source (see Figure~\ref{fig:snj}). %
No structure of the supernova remnant is visible whatsoever since the data is not calibrated yet. %
This illustrates that \textsc{resolve} is able to operate on raw (but already flagged) visibilities that have not been processed further. %
Table~\ref{tab:sn} summarizes the prior parameters for the following reconstruction. %

\begin{table}
  \centering
  \begin{tabular}{ccccccccc}
    &$a$ & $t_0$ & $\bar m$ & $\bar y$ & $\sigma_m$ & $\sigma_{y_0}$ & $\alpha$ & $\beta$ \\
    \toprule
    $A$ & 2 & 2 & $-4$ & 2 & 1 & 2 & 4 & 1\\
    $\lambda$ & 1.5 & 1 & $-4$ & $-37$ & $0.5$ & 1 & 2 & 20\\
    $\phi$ & 1.5 & 1 & $-4$ & $-36$ & $0.5$ & 1 & 2 & 20\\
    \bottomrule
  \end{tabular}
  \caption{SN1006: Prior parameters.}
  \label{tab:sn}
\end{table}

\begin{figure*}
  \centering
  \includegraphics{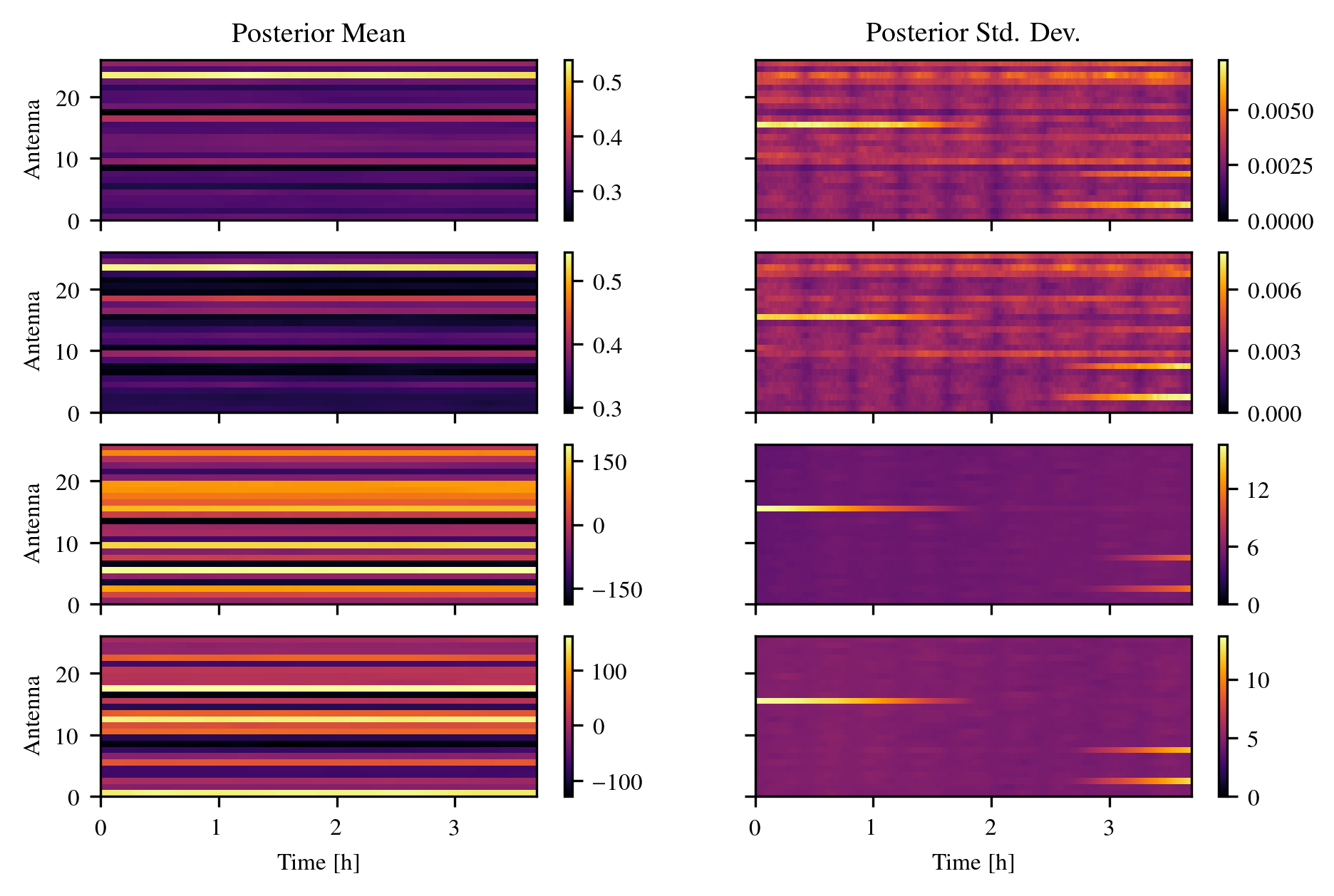}
  \caption{SN1006: Overview over calibration solutions. The four rows indicate amplitude and phase solutions for LL polarization and RR polarization as in Figure~\ref{fig:compcalib}.}
  \label{fig:sncalib}
\end{figure*}

\begin{figure}
  \centering
  \includegraphics{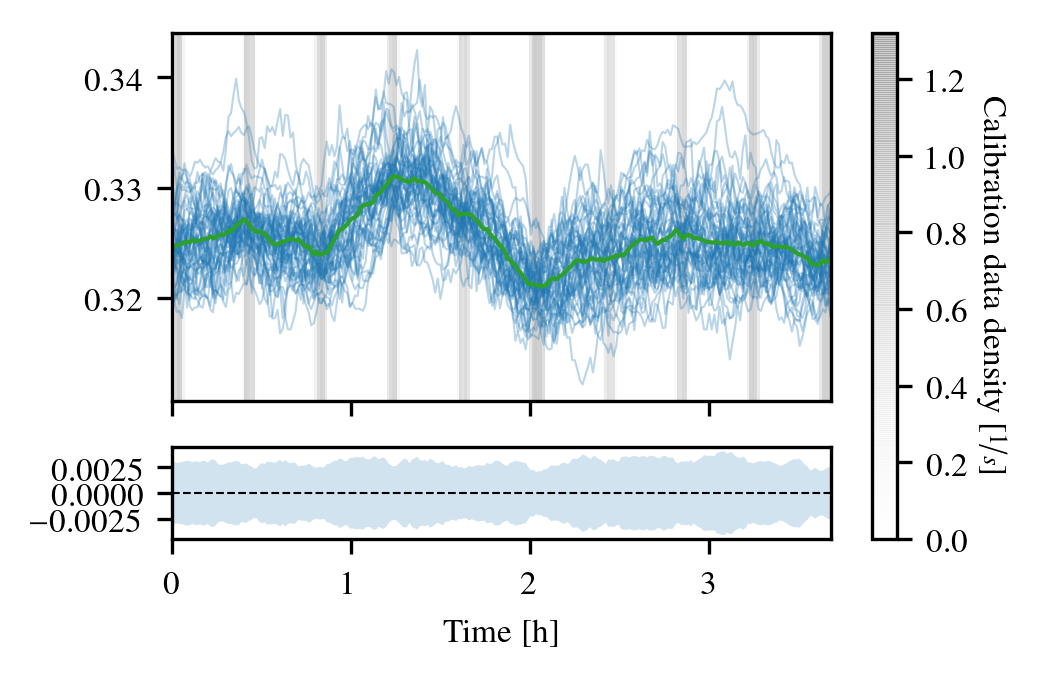}
  \caption{SN1006: Exemplary amplitude solution (see Figure~\ref{fig:exampl}).}
  \label{fig:snampl}
\end{figure}

\begin{figure}
  \centering
  \includegraphics{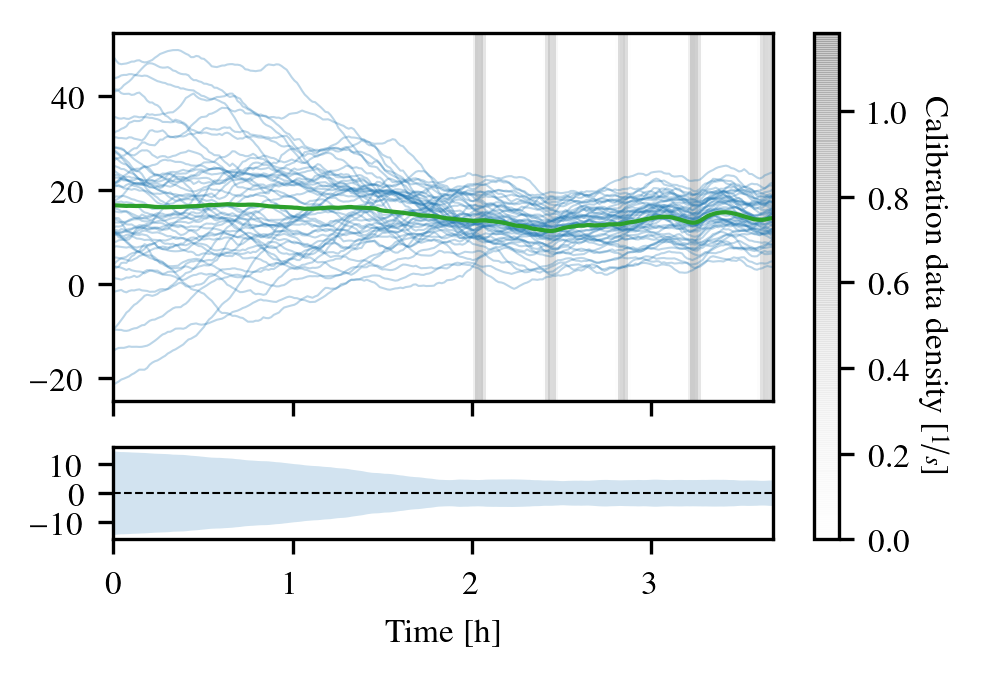}
  \caption{SN1006: Exemplary phase solution (see Figure~\ref{fig:exph}).}
  \label{fig:snphase}
\end{figure}

All calibration solutions are shown in Figure~\ref{fig:sncalib} together with two exemplary plots in Figures~\ref{fig:snampl} and \ref{fig:snphase}. %
The major characteristic of these solutions are hidden in the right-hand column of Figure~\ref{fig:sncalib}: the uncertainty on the calibration decreases whenever the calibrator source is observed as expected. %
Additionally, the uncertainty increases dramatically where the data has been flagged. %
The amplitude solutions are surprisingly stable over time although the prior would allow for more variance in the solution as can be seen from Section~\ref{sec:verification}, where the same prior parameters have been used. %

There is a systematic difference between the samples for the amplitude solutions and those for the phases. %
The former vary around a mean solution whereas the latter exhibit a certain global offset. %
This is explained by the fact that the likelihood is invariant under a pixel-wise global phase shift, which is broken by the prior to a global phase shift to all temporal pixels at once. %
This residual symmetry is again broken by the prior on the zero-mode variance of the phase solutions. %
However, this prior is very weak to allow for phase solutions of arbitrary magnitude. %
Therefore, the phase solutions cannot have an arbitrarily large offset but still can globally vary to some degree, which is shown in Figure~\ref{fig:snphase}.

\begin{figure}
  \centering
  \includegraphics{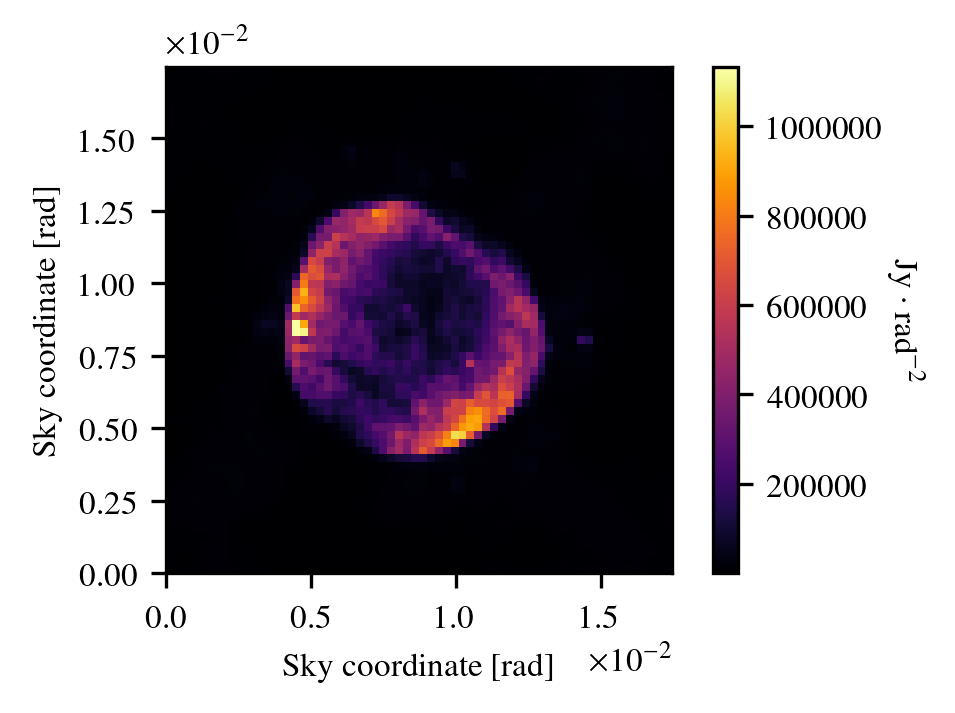}
  \caption{SN1006: Sampled posterior mean of sky brightness distribution.}
  \label{fig:snskymean}
\end{figure}

\begin{figure}
  \centering
  \includegraphics{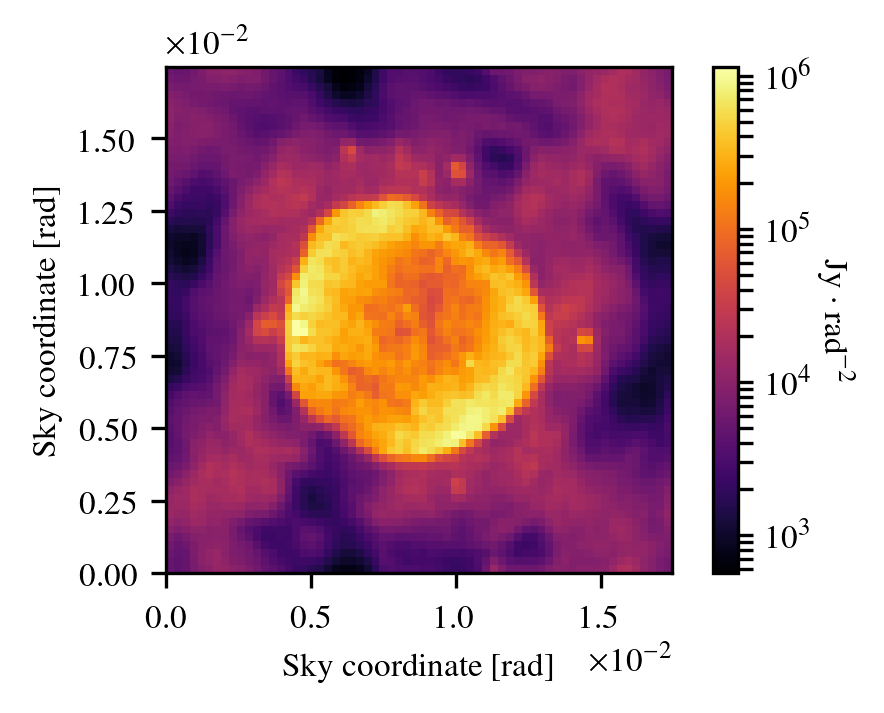}
  \caption{SN1006: Sampled posterior mean of sky brightness distribution (logarithmic color bar).}
  \label{fig:snlogskymean}
\end{figure}

\begin{figure}
  \centering
  \includegraphics{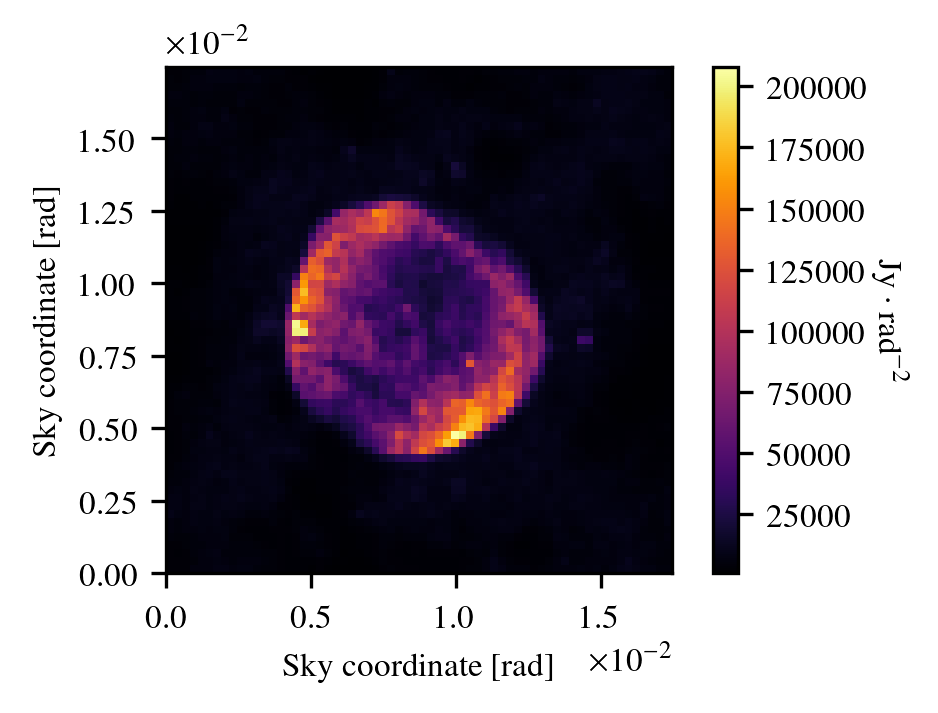}
  \caption{SN1006: Pixel-wise posterior standard deviation of sky brightness distribution.}
  \label{fig:snsd}
\end{figure}

\begin{figure}
  \centering
  \includegraphics{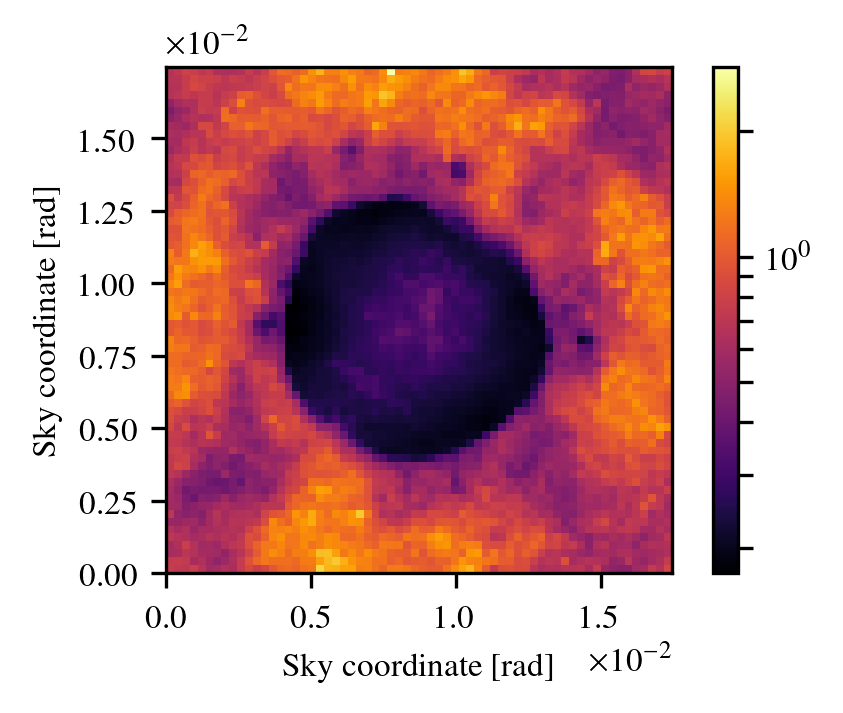}
  \caption{SN1006: Pixel-wise posterior standard deviation normalized by posterior mean of sky brightness distribution.}
  \label{fig:snrelsd}
\end{figure}

Next, the posterior sky brightness is discussed. %
Figures~\ref{fig:snskymean} and \ref{fig:snlogskymean}, along with Figures~\ref{fig:snsd} and \ref{fig:snrelsd}, show the posterior mean and pixel-wise standard deviation of $b(l,m)$. %
The posterior standard deviation is higher wherever more flux is detected. %
Therefore, Figure~\ref{fig:snrelsd} provides a descriptive visualization of the posterior uncertainty of the sky brightness distribution. %

\begin{figure}
  \centering
  \includegraphics{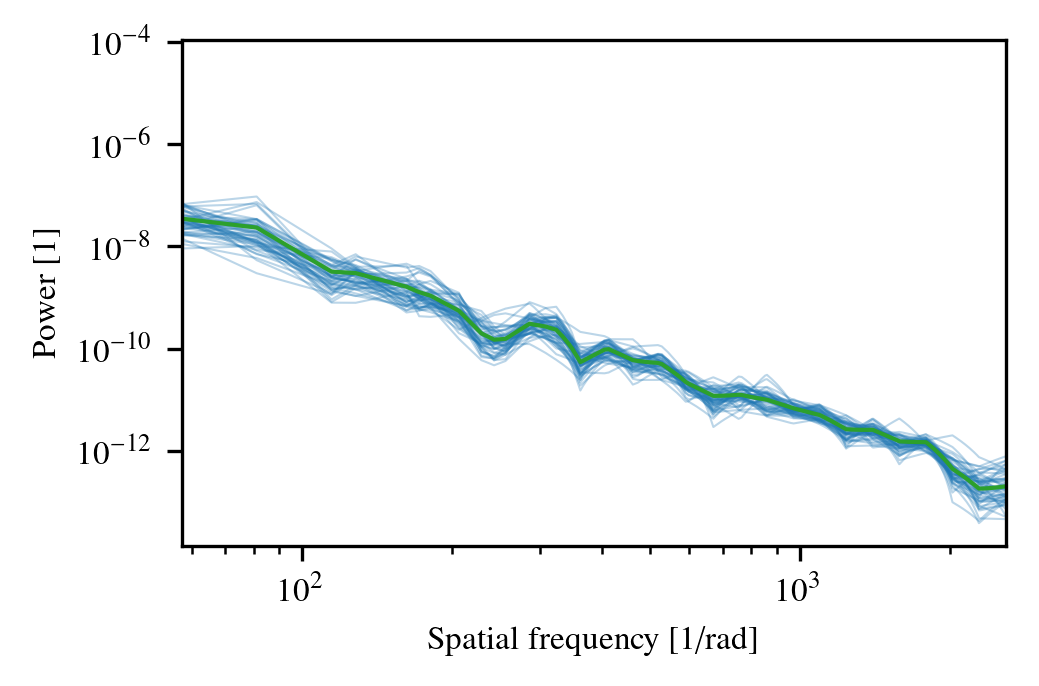}
  \caption{SN1006: Posterior power spectrum of logarithmic sky brightness distribution.}
  \label{fig:snpspec}
\end{figure}

Last but not least the power spectrum of the logarithmic sky brightness distribution also needs to be reconstructed; this is shown in Figure~\ref{fig:snpspec}. %
The power spectrum is more constrained compared to that of section~\ref{sec:verification} since the noise level is much lower in this data set as compared to the synthetic data set. %
We might expect the posterior power spectrum to feature nodes or distinct minima because the Fourier transform of compact objects typically exhibit such. %
This is suppressed by the smoothness prior on the power spectrum. %
However, we note that this does not mean that the algorithm cannot reconstruct the object because it can still choose to not excite the respective modes in $\xi_B$. %

All in all, this demonstrates that \textsc{resolve} is not only able to operate on synthetic data but is actually capable of solving for the sky brightness distribution and the calibration terms at the same time for real data sets. %

\section{Performance and scalability}
Performance and scalability are crucial aspects of the applicability of algorithms. %
The expensive part of the evaluation of the sky model is a fast Fourier transform (FFT), which is in $\mathcal O(n\log n)$ where $\mathrm{n}$ is the total number of pixels of the sky model. %
For real-world data sets the cost for the (de)gridding exceeds the FFTs by far such that one likelihood evaluation is in $\mathcal O(N)$, where $N$ is the number of data points that need to be degridded once for each polarization. %
To compute the sampled KL divergence we need to compute the likelihood $n_s$ times, where $n_s$ is the number of samples (typically 3 -- 20). %
The memory consumption scales linearly with the number of samples used to approximate the KL divergence, number of pixels, and number of data points. %
This is possible since NIFTy is designed such that no explicit matrices need to be stored. %

Both reconstructions in this paper each took $\approx 60$ minutes to be computed on a mobile CPU (Intel(R) Core(TM) i5-4258U CPU @ 2.40GHz) with 4GB main memory. %
The response and adjoint needed to be called $\approx 30000$ times, respectively. %

These values might improve in the future. %
\citet{barnett2018parallel} have proposed a novel gridding kernel that features speedups of several times in first experiments. %
This is possible since it needs relatively small support and can be computed on the fly. %
Also, the structure of the algorithm allows for various forms of parallelization. %
The gridding/degridding can be computed in parallel with OpenMP. %
Moreover, the data set could be split into several parts and distributed on a cluster. %
This is a general feature of Bayesian statistics: a likelihood can be split into the product of two likelihoods each of which contains only a subset of the data. %
Additionally, the evaluation of the KL divergence, which is a sum of few but expensive independent summands, can be distributed. %
Finally, NIFTy offers the (experimental) feature to distribute large fields on a cluster. %
Orthogonal to computational speedup ideas the algorithm might also benefit from compressing the likelihood itself such that fewer (de)gridder calls are necessary. %

\section{Conclusions}
\label{sec:conclusions}
We have presented the probabilistic \textsc{resolve} algorithm for simultaneous calibration and imaging. %
After a derivation from first principles of the full posterior probability distribution for the joint calibration and imaging algorithm \textsc{resolve}, it has been shown how this distribution can be approximated by a multivariate Gaussian probability distribution to render the problem computationally solvable. %
This method is called MGVI and provides a prescription for how to draw samples from the approximate posterior distribution. %
The calibration algorithm \textsc{resolve} has been verified on synthetic data. %
The results indicate that the uncertainty quantification is qualitatively sensible but should be taken with a grain of salt since MGVI systematically underestimates posterior variance. %
Furthermore, it has been demonstrated that the algorithm has the capability to reconstruct a sky brightness distribution of a intricate source, the supernova remnant SN1006, together with uncertainty information from raw VLA L-band data. %

There are many open ends to continue the investigation that we started with this paper. %
First, the model for the sky brightness distribution may include point source and multifrequency correlations. %
On top of that the response may be described more thoroughly. %
Direction-dependent calibration and nontrivial primary beam effects may be taken into account. %
Moreover, we performed the flagging by a standard CASA flagging algorithm. %
This can be replaced with an algorithm rooted in information theory that unifies flagging with calibration/imaging. %
Additionally, a major/minor cycle scheme similar to that in CLEAN may be introduced to avoid to frequent (de)gridding operations. %
This is necessary to apply \textsc{resolve} to big data sets from telescopes such as MeerKAT. %
Finally, \textsc{resolve} can be extended to polarization imaging. On an orthogonal track \textsc{resolve} may be used for imaging of a variety of sources from different telescopes including ALMA and especially the Event Horizon Telescope. %

\begin{acknowledgements}
  We would like to thank Jamie Farnes for his workshop on calibration with CASA at the Power of Faraday Tomography 2018 conference and his script for calibrating the data set at hand with CASA, which enabled development of the IFT calibration algorithm. We thank Landman Bester, Ben Hugo, Jakob Knollmüller, Julian Rüstig, Oleg Smirnov, and Margret Westerkamp for numerous discussions, explanations, and feedback, and Martin Reinecke for his work on \textsc{NIFTy}, which was the crucial technical premise of the project. We acknowledge financial support by the German Federal Ministry of Education and Research (BMBF) under grant 05A17PB1 (Verbundprojekt D-MeerKAT). %
We thank the anonymous referee for insightful comments and the language editor for substantially improving the text quality. %
\end{acknowledgements}

\bibliographystyle{aa}
\bibliography{bib.bib}

\end{document}